\documentclass[%
reprint,
superscriptaddress,
nofootinbib,
 amsmath,amssymb,
 aps,
]{revtex4-2}

\usepackage{graphics,epsfig,psfrag}
\usepackage{color}
\usepackage{xcolor}
\usepackage{graphicx}
\usepackage{dcolumn}
\usepackage{bm}
\usepackage{hyperref}

\usepackage{mathrsfs}

\begin{document}

\title{Light propagation in a plasma on Kerr spacetime.\\
II. Plasma imprint on photon orbits}

\author{Volker Perlick}
\email{perlick@zarm.uni-bremen.de}
\affiliation{ZARM, University of Bremen, 28359 Bremen, Germany}

\author{Oleg Yu. Tsupko}
\email{tsupkooleg@gmail.com; tsupko@cosmos.ru}
\affiliation{ZARM, University of Bremen, 28359 Bremen, Germany}
\affiliation{Space Research Institute of Russian Academy of Sciences, Profsoyuznaya 84/32, Moscow 117997, Russia}

\date{\today}

\begin{abstract}
In this paper, light propagation in a pressure-free non-magnetized plasma on Kerr spacetime is considered, which is a continuation of our previous study [Phys. Rev. D 95, 104003 (2017)]. It is assumed throughout that the plasma density is of the form that allows for the separability of the Hamilton-Jacobi equation for light rays, i.e., for the existence of a Carter constant. Here we focus on the analysis of different types of orbits and find several peculiar phenomena which do not exist in the vacuum case. We start with studying  spherical orbits, which are contained in a coordinate sphere $r = \mathrm{constant}$, and conical orbits, which are contained in a coordinate cone $\vartheta = \mathrm{constant}$. In particular, it is revealed that in the ergoregion in the presence of a plasma there can exist two different spherical light rays propagating through the same point. Then we study circular orbits and demonstrate that, contrary to the vacuum case, circular orbits can exist off the equatorial plane in the domain of outer communication of a Kerr black hole. Necessary and sufficient conditions for that are formulated. We also find a compact equation for circular orbits in the equatorial plane of the Kerr metric, with several examples developed. Considering the light deflection in the equatorial plane, we derive a new exact formula for the deflection angle which has the advantage of being directly applicable to light rays both inside and outside of the ergoregion. Remarkably, the possibility of a non-monotonic behavior of the deflection angle as a function of the impact parameter is demonstrated in the presence of a non-homogeneous plasma. Furthermore, in order to separate the effects of the black-hole spin  from the effects of the plasma, we investigate weak deflection gravitational lensing. We also add some further comments to our discussion of the black-hole shadow which was the main topic of our previous paper.
\end{abstract}

\maketitle



\section{Introduction}

If a gravitating body is surrounded by a medium, the trajectories of light rays are influenced not only by the gravitational field but also by the medium. In realistic astrophysical applications the medium is a plasma which is a dispersive medium, i.e., the influence on light rays depends on the photon frequency: Whereas this influence is usually negligible for optical and higher frequencies, it may be measurable for radio frequencies if the plasma density is sufficiently high. This may be the case, in particular, in the vicinity of black holes which leads to a complex interplay of gravitation and dispersive refraction.

The Hamilton formalism for light rays in a curved spacetime in the presence of a dispersive medium is detailed in a book by Synge \cite{Synge-1960}. Although not explicitly mentioned by Synge, his equations include the case of light propagation in a non-magnetized pressure-free plasma, see also \cite{Bicak-Hadrava-1975, Kichenassamy-Krikorian-1985, Krikorian-1999, Kulsrud-Loeb-1992}. For the plasma case, a rigorous derivation of this Hamilton formalism from Maxwell's equations on a general-relativistic spacetime was given by Breuer and Ehlers \cite{Breuer-Ehlers-1980, Breuer-Ehlers-1981, Breuer-Ehlers-1981-AA} who even allowed for a magnetized plasma. In view of applications to astrophysics, a magnetized plasma was also considered by Broderick and Blandford \cite{Broderick-Blandford-2003, Broderick-Blandford-2003-ASS, Broderick-Blandford-2004}.

If the gravitational field is weak and the deflection angle of light is small, the effect of the plasma can be separated from the effect of gravity. This case was first investigated, for a non-magnetized plasma, by Muhleman et al \cite{Muhleman-1970} who considered the deflection of light by the Solar corona, see also \cite{Muhleman-1966, Lightman-1979}. In the same approximation, Bliokh and Minakov \cite{Bliokh-Minakov-1989} studied the properties of lensed images when a gravitational lens is surrounded by a plasma. A detailed analytical description of gravitational lensing in the presence of a plasma under the weak-deflection assumption was presented in a series of papers by Bisnovatyi-Kogan and Tsupko \cite{BK-Tsupko-2009, BK-Tsupko-2010, Tsupko-BK-2013}. On the basis of Synge's approach, they calculated the deflection angle in different scenarios, including the correction to the vacuum angle due to a homogeneous plasma. As a lens, a Schwarzschild black hole and different extended spherically symmetric mass distributions were considered. The images are predicted to blur into a 'rainbow', since different deflection angles correspond to photons of different frequencies. The study of Kerr lensing in a plasma in the weak-deflection case was performed by Morozova et al. \cite{Morozova-2013} who considered a slowly rotating black hole and a homogeneous plasma distribution. In a series of papers by Crisnejo et al \cite{Crisnejo-Gallo-2018, Crisnejo-Gallo-Rogers-2019, Crisnejo-Gallo-Villanueva-2019, Crisnejo-Gallo-Jusufi-2019}, higher-order terms of weak deflection in the presence of a plasma were found for different spacetimes. In particular, Schwarzschild and Kerr black holes were considered.

Without the weak-field approximation the combined action of gravity and a plasma on light rays was first studied by Perlick \cite{Perlick-2000} who calculated the ray deflection in the Schwarzschild metric and in the equatorial plane of the Kerr metric in the presence of a plasma. Properties of higher-order images in the case of gravitational lensing by a Schwarzschild black hole in the presence of a plasma were investigated analytically by Tsupko and Bisnovatyi-Kogan \cite{Tsupko-BK-2013}. In that paper, the strong deflection limit for the case of a homogeneous plasma was derived. Moreover, ray trajectories near compact objects (described by the Schwarzschild metric) surrounded by a plasma were studied in a series of works by Rogers \cite{Rogers-2015, Rogers-2017-MG14, Rogers-2017a, Rogers-2017b}. In particular, different power-law plasma distributions were analyzed. For a discussion of the Kerr case beyond the weak-deflection approximation see also Crisnejo, Gallo and Jusufi \cite{Crisnejo-Gallo-Jusufi-2019}.

In recent years, the interest in gravitational lensing in the presence of a plasma increased considerably. There are studies of strong lens systems with multiple images \cite{Er-Mao-2014, Er-Mao-2022, Kumar-Beniamini-2023, Crisnejo-Gallo-2023-perturbative, BK-Tsupko-2023-time-delay}, microlensing \cite{Tsupko-BK-2020-microlensing, Sun-Er-Tsupko-2023}, spatial dispersion in Kerr spacetime \cite{Kimpson-2019a}, wave effects for propagation near the Sun \cite{Turyshev-2019a, Turyshev-2019b}, time delay \cite{Muhleman-1966, Er-Mao-2014, Er-Yang-Rogers-2020, Er-Mao-2022, Kumar-Beniamini-2023, BK-Tsupko-2023-time-delay}. For some other related studies see \cite{Schulze-Koops-Perlick-Schwarz-2017, Sareny-2019,  Kimpson-2019b, Wagner-Er-2020, Matsuno-2021, Briozzo-Gallo-2023, Chainakun-2022, Guerrieri-Novello-2022}.

In particular, the influence of a (non-magnetized) plasma on the shadow of black holes has been studied by many authors. In \cite{Perlick-Tsupko-BK-2015}, the shadow of spherically symmetric black holes (and other compact objects) surrounded by a plasma was calculated analytically. In our previous work \cite{Perlick-Tsupko-2017} (hereinafter Paper I) we investigated the propagation of light rays in the Kerr metric in the presence of a plasma without approximation. The equations of motion were derived for a photon that moves anywhere in the domain of outer communication under an arbitrary inclination in a Kerr spacetime with arbitrary spin parameter. One of the main results of this work was that the Hamilton-Jacobi equation for the light rays can be separated only for some class of plasma distributions. Accordingly, for such plasma distributions the equations of motion are reduced to first-order differential equations, which greatly simplifies further calculations. In particular, for this class of plasma distributions we could analytically calculate the size and the shape of the black-hole shadow for an arbitrary position of the observer in the domain of outer communication. On the basis of these analytical results we constructed pictures of the shadow for various plasma distributions.

These works were followed by many papers by different authors who considered the shadow in the presence of a plasma, see e.g. \cite{Yan-2018, Chowdhuri-2021-shadow-expand, Badia-Eiroa-2021, Badia-Eiroa-2023, Li-2022, Bezdekova-2022, Briozzo-Gallo-Madler-2023}. If the plasma distribution around a Kerr black hole does not satisfy the separability condition, the black-hole shadow can only be calculated by a numerical integration of the equations of motion which has been done for some examples \cite{Huang-2018, Zhang-2023}. For the shadow of a spherically symmetric black hole surrounded by a medium with an arbitrary frequency-dependent index of refraction see \cite{Tsupko-2021}.

In this paper, we continue our investigation of ray propagation in a plasma on Kerr spacetime started in Paper I (Perlick and Tsupko \cite{Perlick-Tsupko-2017}). Here we focus mainly on the analysis of different types of orbits, filling some gaps that have been left in the previous literature. As a result, we reveal several peculiar phenomena which do not exist in the vacuum case. We also add some comments on the shadow and we discuss in some detail the deflection angle of light, both exact and in the weak-deflection approximation.

The paper is organized as follows. Starting out from the equations of motion for light rays (Section \ref{sec:geodesics-general}) on the Kerr spacetime in a plasma that satisfies the separability condition, we consider spherical and conical orbits (Section \ref{sec:spherical-conical}). Further, we investigate in detail the circular orbits on and off the equatorial plane, for an arbitrary plasma distribution that satisfies the separability condition and for a number of particular examples (Section \ref{sec:circular}). In Section \ref{sec:shadow} we comment on the shadow, discussing in particular the observability or non-observability of ``fishtails'' in the boundary curve of the shadow found in Paper I. After specifying the equations of motion for light rays in the equatorial plane, we derive an exact formula for the deflection angle which is suitable for all rays that come in from infinity and return to infinity, including those that enter into the ergoregion (Section \ref{sec:exact}). Up to this point no slow-rotation or weak-deflection approximation was involved. In the remaining two sections we still allow for an arbitrary black-hole spin but we assume weak deflection. In Section \ref{sec:weak} we derive the deflection angle for a light ray in the equatorial plane in this approximation.
In Section \ref{sec:images} we solve the lens equation for this case and discuss the influence of the black-hole spin and of the plasma on the image positions. 
Then we end with some Conclusions.

\section{Equations of motion for light rays} \label{sec:geodesics-general}

Recall that the propagation of light rays in a non-magnetized pressure-less plasma on a spacetime with metric $g_{\mu \nu} dx^{\mu} dx^{\nu}$ can be described by the Hamiltonian (\cite{Breuer-Ehlers-1980}, also see \cite{Perlick-2000, BK-Tsupko-2009, BK-Tsupko-2010})
\begin{equation}\label{eq:Hg}
H (x, p) = \dfrac{1}{2} \big( g^{\mu \nu} (x) p_{\mu} p_{\nu} 
+ \omega _p (x) ^2 \big)
\, .
\end{equation}
Here $x = (x^0,x^1,x^2,x^3)$ are the spacetime coordinates, 
$p=(p_0,p_1,p_2,p_3)$ are the canonical momentum coordinates, 
and 
\begin{equation}
\omega_p(x) ^2 = \frac{4 \pi e^2}{m_e} N(x)
\end{equation}
is the plasma frequency, where $e$ and $m_e$ are the electron charge and mass, respectively, and $N$ is the electron number density. The light rays are the solutions to Hamilton's equations
\begin{equation}\label{eq:Ham}
\frac{dx^{\mu}}{ds} = \frac{\partial H}{\partial p_{\mu}} \, , \quad
\frac{dp_{\mu}}{ds} = - \, \frac{\partial H}{\partial x^{\mu}} \, , \quad
H(x,p) = 0 \, ,
\end{equation}
where $s$ is a curve parameter.

In this paper we assume that the spacetime metric is the Kerr metric. In Boyer-Lindquist coordinates $x=(t, r, \vartheta , \varphi )$, it reads
\begin{equation} \label{Kerr-metric}
g _{\mu \nu} dx^{\mu} dx^{\nu}  =  
 -c^2  \left(1-\frac{2mr}{\rho^2}\right) dt^2 + 
\frac{\rho^2}{\Delta} dr^2 + \rho^2 d\vartheta^2 
\end{equation}
\[
+ \, \mathrm{sin}^2\vartheta\left(r^2+a^2+
\frac{2mra^2\mathrm{sin}^2\vartheta}{\rho^2}\right)
d\varphi^2 - \, \frac{4mra \mathrm{sin} ^2\vartheta}{\rho^2} \, c \, dt \, d\varphi
\]
where
\begin{equation}\label{eq:Deltarho}
\Delta = r^2 + a^2 - 2mr \, , \quad
\rho ^2 = r^2 + a^2 \mathrm{cos} ^2 \vartheta \, .
\end{equation}
Here $m$ is the mass parameter and $a$ is the spin parameter,
\begin{equation}\label{eq:mM}
m = \dfrac{GM}{c^2} \, , \quad a = \frac{J}{Mc} \, ,
\end{equation}
where $M$ is the mass and $J$ is the angular momentum of the black hole. The Kerr parameter $a$ is restricted to the interval $0 \le a \le m$. Here $0 \le a$ can be assumed without loss of generality because we are free to make a coordinate transformation $\varphi \mapsto - \varphi$ which has the same effect as changing the sign of $a$. The assumption $a \le m$ restricts us to the case of a black hole, as opposed to a naked singularity. In the following we want to discuss effects that are observable for an observer who is outside of the black hole. Therefore, we restrict to the domain of outer communication, i.e., to the domain outside of the black-hole horizon which is at $r = m + \sqrt{m^2-a^2}$.

The index of refraction of a plasma under consideration is 
\begin{equation}\label{eq:n}
n \big( x, \omega (x) \big) ^2 = 1 - \frac{\omega _p (x)^2}{\omega (x) ^2}
\, ,
\end{equation}
where 
\begin{equation}
\omega (x) = 
\frac{\omega _0}{\sqrt{1- \dfrac{2 \, m \, r}{\rho^2 }}}
\label{eq:omegax}
\end{equation}
is the photon frequency measured at $x$ by an observer on a $t$-line.
Here and in the following,
\begin{equation}
    \omega _0 = -p_t / c \, ,
\end{equation}
where we have chosen the sign such that $\omega _0$ is positive for future-oriented light rays outside of the ergoregion. Inside the ergoregion (\ref{eq:omegax}) does not make sense because the expression under the square-root becomes negative; this reflects the fact that there the $t$-lines are spacelike. Note that, by (\ref{eq:omegax}), the constant of motion $\omega _0$ is to be interpreted as the frequency at infinity. Of course, this interpretation makes sense only for those light rays that actually reach infinity.  \\

In Paper I we have found the necessary and sufficient condition for separability of the Hamilton-Jacobi equation, i.e., for the existence of a generalized Carter constant (see also the discussion in the Introduction). We have shown that separability requires the plasma frequency to be of the form
\begin{equation}\label{eq:sepcon}
\omega_p (r , \vartheta ) ^2 =
\frac{f_r(r)+f_{\vartheta} ( \vartheta )}{r^2 + a^2 \mathrm{cos} ^2 \vartheta}
\end{equation}
where $f_r(r)$ is an arbitrary function of $r$ and $f_{\vartheta} ( \vartheta )$ is an arbitrary function of $\vartheta$.

If the separability condition (\ref{eq:sepcon}) is satisfied, the equations of motion for light rays have the following form (Paper I):
\begin{equation}\label{eq:dott}
\rho ^2 \,  \dot{t} =
\dfrac{ \Big(  (r^2+a^2) \rho ^2 + 2mra^2 \mathrm{sin}^2 \vartheta \Big) 
\omega _0 - 2mra p_{\varphi}}{c \, \Delta }
 \, ,
\end{equation}
\begin{equation}\label{eq:dotphi}
\rho ^2 \dot{\varphi}  =
\dfrac{2mra \, \mathrm{sin} ^2 \vartheta  \, \omega _0
+ \big( \rho^2 - 2mr \big) p_{\varphi}}{
\Delta \, \mathrm{sin}^2 \vartheta} \, , 
\end{equation}
\begin{equation}\label{eq:dottheta}
\rho ^4 \dot{ \vartheta}{}^2 = K 
-\Big( \dfrac{p_{\varphi}}{\mathrm{sin} \, \vartheta} -
a \mathrm{sin} \, \vartheta \, \omega _0 \Big)^2
- f_{\vartheta} ( \vartheta ) 
\, ,  
\end{equation}
\begin{equation}\label{eq:dotr}
\rho ^4 \dot{r}{}^2= - K \Delta  +
 \Big( (r^2+a^2) \, \omega _0 - a p_{\varphi} \Big) ^2
- f_r (r) \, \Delta 
\, . 
\end{equation}
Here and in the following the overdot denotes the derivative with respect to the curve parameter $s$ from (\ref{eq:Ham}).
Note that we have changed the sign of $\omega _0$ in comparison to Paper I. The reason is that here we want to have $\omega _0$ positive for future-oriented rays, as mentioned above. In Paper I we found it convenient to have $\omega _0$ positive for past-oriented rays because the main focus was on the calculation of the shadow which is usually done by considering light rays that issue from the observer position into the past. In the present paper, however, we also want to discuss the deflection angle and for this purpose it would not be natural to consider past-oriented rays. \\

In order to solve the equations of motion (\ref{eq:dott}),(\ref{eq:dotphi}),(\ref{eq:dottheta}),(\ref{eq:dotr}), one has to specify the black hole parameters $m$ and $a$, and choose the functions characterizing the plasma distribution $f_r(r)$ and $f_{\vartheta} ( \vartheta )$. As usual, it is convenient to express all lengths in units of $m$. Additionally, one has to specify several constants of motion on the right-hand side of these equations: the photon frequency at infinity $\omega_0$, the momentum component $p_\varphi$ and the generalized Carter constant $K$.

It is important to emphasize that not all values of the constants of motion are possible.
As the right-hand side of (\ref{eq:n}) must be non-negative, Eqs.(\ref{eq:sepcon}) and (\ref{eq:omegax}) require
\begin{equation}
\dfrac{
(\rho ^2-2mr ) \big( f_r(r)+f_{\vartheta} (\vartheta ) \big)
}{
\rho ^4 \omega _0 ^2 
}
\le  1 \, . 
\label{eq:omega0con}
\end{equation}
This condition restricts the possible values of $\omega _0$ outside of the ergoregion. Inside the ergoregion, the first bracket in the numerator becomes negative, therefore the condition (\ref{eq:omega0con}) is satisfied for all $\omega _0 \neq 0$.

As we restrict to the domain of outer communication, where $\Delta > 0$, it is obvious that $(\rho ^4 / \omega _0 ^2 ) \big( \dot{\vartheta}{}^2 + \dot{r}{}^2 / \Delta \big) \ge 0$. By (\ref{eq:dottheta}) and (\ref{eq:dotr}), after some elementary calculation, this inequality reads 
\[
\Big( 
(\rho ^2 - 2mr ) \dfrac{p_{\varphi}}{\omega _0} + 2mra \, \mathrm{sin} ^2 \vartheta 
\Big) ^2 
\le
\]
\begin{equation}
\Delta \, \rho ^4 \mathrm{sin} ^2 \vartheta 
\Big( 1 - 
\dfrac{
(\rho ^2 - 2mr ) \big( f_r (r) + f_{\vartheta} (\vartheta ) \big)
}{
\rho ^4 \omega _0 ^2
}
\Big) \, .
\label{eq:pphicond}
\end{equation}
Note that, by (\ref{eq:omega0con}), the right-hand side of this inequality is non-negative for all allowed values of $\omega _0$. For each $r$, $\vartheta$ and $\omega _0$, condition (\ref{eq:pphicond}) restricts the possible values of $p_{\varphi}$.

\section{Spherical and conical light rays}
\label{sec:spherical-conical}

Of particular interest are \emph{spherical light rays}, which are contained in a coordinate sphere $r = \mathrm{const.}$, and \emph{conical light rays}, which are contained in a coordinate cone $\vartheta = \mathrm{const.}$ In this Section, we study these types of orbits.

The relevant equations for spherical light rays have already been obtained in Paper I but we want to rewrite them here in a different form that is more convenient. As spherical light rays have to satisfy $\dot{r}=0$ and $\ddot{r}=0$, one finds from (\ref{eq:dotr}) that their constants of motion $p _{\varphi}/\omega _0$ and $K/\omega _0^2$ have to satisfy
\begin{equation}
\dfrac{K}{\omega _0 ^2} - \dfrac{1}{\Delta}
\Big( r^2 +a^2 - \dfrac{a p _{\varphi}}{\omega _0} \Big) ^2
+ \dfrac{f_r(r)}{\omega _0 ^2} = 0 \, ,
\label{eq:Ksph}
\end{equation}
\[
\Big( r^2 +a^2 - \dfrac{a p _{\varphi}}{\omega _0} \Big) ^2
- \dfrac{2 r \Delta}{(r-m)} 
\Big( r^2 +a^2 - \dfrac{a p _{\varphi}}{\omega _0} \Big) 
\]
\begin{equation}
+ \, \dfrac{\Delta ^2 f_r ' (r)}{2 (r-m) \omega _0 ^2} = 0 \, .
\label{eq:pphispher}
\end{equation}
Equation (\ref{eq:pphispher}) is found by taking the derivative of (\ref{eq:dotr}) with respect to $s$ and then dividing both sides by $\dot{r}$. By continuity, the resulting equation is true also at points where $\dot{r}=0$.

Solving the quadratic equation (\ref{eq:pphispher}) for $a p_{\varphi} / \omega _0$ gives two solutions,
\begin{equation}\label{eq:pphisph}
a \, \dfrac{p_{\varphi}}{\omega_0} = r^2+a^2- \dfrac{r \Delta }{r-m} 
\left(
1 \pm \sqrt{1- \dfrac{f_r'(r) (r-m)}{2 \omega _0^2 r^2}} \right) \, .
\end{equation}

From this equation we read that, for every given $\omega _0$, spherical light rays can exist at a point with radial coordinate $r$ only if 
\begin{equation}
    \dfrac{f_r'(r) (r-m)}{2 \omega _0^2 r^2} \le 1 
\end{equation}
and that there are at most two such light rays. With $a p_{\varphi} / \omega _0$ determined, Eq.(\ref{eq:Ksph}) then gives the corresponding value of the Carter constant $K$. If such a spherical light ray actually exists is determined by the inequality (\ref{eq:pphicond}). If we multiply this inequality by $a^2$ and insert (\ref{eq:pphisph}), we find after some elementary algebra:
\[
- \, r^2 \Delta (\rho ^2-2mr) 
\sqrt{1- \dfrac{f_r'(r) (r-m)}{2 \omega _0^2 r^2}}^{\, 2}
\]
\[
\pm \,
2mr \Delta (r^2-a^2 \mathrm{cos} ^2 \vartheta )
\sqrt{1- \dfrac{f_r'(r) (r-m)}{2 \omega _0^2 r^2}}
\]
\[    
\ge
(m^2-a^2) \rho ^4 + ( \rho ^2 + 2 m r ) \Delta \, a^2 \mathrm{cos}^2 \vartheta
\]
\begin{equation}
+ \, (r-m)^2 a^2 \mathrm{sin} ^2 \vartheta \,
\dfrac{f_r(r)+f_{\vartheta} (\vartheta )}{\omega ^2} \, .
\label{eq:photonreg}
\end{equation}
The inequality (\ref{eq:photonreg}) determines the \emph{photon region}, i.e., the set of all points through which spherical light rays exist. Note that in vacuum condition (\ref{eq:photonreg}) can hold only with the plus sign. In a plasma, however, it is possible that  it holds for some coordinates $r$ and $\vartheta$ with both signs. This means that there are two spherical light rays through a point with these coordinates; for one ray the constant of motion $p _{\varphi} /\omega _0$ is given by (\ref{eq:pphisph}) with the plus sign and for the other ray with the minus sign.

Condition (\ref{eq:photonreg}) is equivalent to condition (43) of Paper I, but it is more convenient: As we consider the black-hole case ($a^2 \le m^2$) and restrict to the domain of outer communication ($\Delta > 0$), the right-hand side of (\ref{eq:photonreg}) is manifestly positive. Also, these assumptions imply that $\Delta (r^2-a^2 \mathrm{cos}^2 \vartheta )$ is positive. Therefore, if (\ref{eq:photonreg}) holds with the minus sign, then it also holds with the plus sign. As the plus-minus sign in (\ref{eq:photonreg}) corresponds to the plus-minus sign in (\ref{eq:pphisph}), this implies the following: Through each point of the photon region there is either exactly one spherical light ray, with constant of motion $p_{\varphi}$ satisfying (\ref{eq:pphisph}) with the plus sign, or there are two spherical light rays, with $p_{\varphi}$ satisfying (\ref{eq:pphisph}) for the first one with the plus sign and for the second one with the minus sign. Moreover, it is also obvious that (\ref{eq:photonreg}) can hold with the minus sign only if the first term on the left-hand side is positive and dominates the second one. The first condition requires that $(\rho ^2-2mr)$ is negative and the second condition requires that the square-root is bigger than 1. This demonstrates that the case that there are two spherical light rays through a certain point can happen only inside the ergoregion and only if $f_r'(r)$ is negative.

It is also true that (\ref{eq:pphisph}) and, thus (\ref{eq:photonreg}) can hold with the minus sign only if the plasma density is non-decreasing in the radial direction. To prove this, let us assume that the plasma density is decreasing,
\begin{equation}
 \dfrac{\partial}{\partial r} \omega _p (r , \vartheta ) ^2  = 
    \dfrac{\rho ^2 f_r'(r) -2r (f_r(r)+f_{\vartheta} ( \vartheta ) \big)}{\rho^4} < 0 \, ,
\label{eq:decreasing}
\end{equation}
and that (\ref{eq:pphispher}) and, thus, (\ref{eq:photonreg}) hold with the minus sign. We have just seen that the latter condition requires that the square-root in (\ref{eq:pphisph}) and (\ref{eq:photonreg}) is bigger than 1. Therefore, inserting (\ref{eq:pphisph}) into  (\ref{eq:Ksph}) demonstrates that
\begin{equation}
    K + f_r(r) < \dfrac{\Delta f_r'(r)}{2 (r-m)} \, .
\end{equation}
On the other hand, (\ref{eq:dottheta}) requires 
\begin{equation}
    -K + f_{\vartheta} (\vartheta ) \le 0 \, .
\end{equation}
Adding these two inequalities together yields
\begin{equation}
    f_r (r) + f_{\vartheta} (\vartheta ) < \dfrac{\Delta f_r'(r)}{2 (r-m)} \, .
\end{equation}
In combination with our assumption (\ref{eq:decreasing}) this implies
\begin{equation}
\dfrac{(r-m)}{\Delta} \big( f_r(r) + f_{\vartheta} (\vartheta ) \big) <
\dfrac{r}{\rho ^2} \big( f_r(r) + f_{\vartheta} (\vartheta ) \big)
\end{equation}
and thus, as $\big( f_r(r)+f_{\vartheta} (\vartheta ) \big) /(\Delta \rho ^2)>0$,
\begin{equation}
    (r-m) \rho ^2- r \Delta = 
(r-m)(rm+a^2 \mathrm{cos} ^2 \vartheta )+ r (m^2-a^2)  
    < 0 \, .
\end{equation}
As $a^2 \le m^2$ and $m \le r$, according to our general assumptions, this is the desired contradiction, so we have proven that (\ref{eq:pphisph}) and (\ref{eq:photonreg}) can hold with the minus sign only if the plasma density is non-decreasing.

If the horizon is approached, $\Delta \to 0$, the left-hand side of (\ref{eq:photonreg}) goes to 0. The right-hand side, however, is positive and bounded away from zero, unless in the extreme case, $a^2=m^2$, where $\Delta \to 0$ means $r \to m$. So, if we exclude the extreme case, the photon region is always separated from the horizon by a finite interval of the radius coordinate.

In the Schwarzschild case, $a=0$, the inequality in (\ref{eq:photonreg}) reduces to an equality which can hold only with the upper sign,
\begin{equation}
    m = (r-2m) \sqrt{1-f_r'(r) \dfrac{(r-m)}{2 r^2 \omega _0^2}} \, .
\label{eq:SchwSph}
\end{equation}
So in this case the photon region becomes a photon sphere. Note, however, that the light rays that are confined to this photon sphere are not in general circles.

In the extreme case, $a^2=m^2$, inequality (\ref{eq:photonreg}) simplifies to
\begin{gather}\label{eq:photonregext}
- \, r^2 (\rho ^2-2mr )
\sqrt{1-f_r'(r) \dfrac{(r-m)}{2 r^2 \omega _0^2}}^{\, 2}
\\
\nonumber
\pm \, 2mr (r^2-m^2 \mathrm{cos}^2 \vartheta )
\sqrt{1-f_r'(r) \dfrac{(r-m)}{2 r^2 \omega _0^2}}
\\
\nonumber
\ge
m^2 \mathrm{cos} ^2 \vartheta
\big( \rho ^2+2mr \big)
+ \frac{f_r(r)+ f_{\vartheta} ( \vartheta )}{\omega _0 ^2} \,
m^2 \mathrm{sin} ^2 \vartheta 
\, .  
\end{gather} 
To demonstrate that the condition for spherical light rays can indeed hold with the minus sign, we give a specific example: Consider an extreme Kerr black hole with a plasma density that satisfies $f_r(3m/2)+f_{\vartheta} ( \pi /2) = m^2 \omega_0^2$ and $f_r'(3m/2)=-216 \, m \, \omega _0^2$. Then (\ref{eq:photonregext}) is satisfied, with a strict inequality sign, at $r=3m/2$ and $\vartheta = \pi /2$ for both signs; this means that there are two spherical light rays through each point of this circle, and by continuity also through each point of a neighborhood of this circle.\\ 

We now turn to the conical light rays which are determined by the equations $\dot{\vartheta}=0$ and $\ddot{\vartheta}=0$. With the help of (\ref{eq:dottheta}) these two equations can be written as
\begin{equation}\label{eq:Kcon}
\dfrac{K}{\omega _0^2}
=
\Bigg(
\dfrac{p_{\varphi}}{\omega _0 \mathrm{sin} \, \vartheta} - a \, \mathrm{sin} \, \vartheta \Bigg)^2
+
\dfrac{f_\vartheta (\vartheta)}{\omega _0^2}
\, ,
\end{equation} 
\begin{equation}\label{eq:pphicon}
2 \, \dfrac{\mathrm{cos} \, \vartheta}{\mathrm{sin}^3 \vartheta} \Bigg(
\dfrac{p_{\varphi}^2}{\omega _0^2} - a^2 \mathrm{sin}^4 \vartheta \Bigg)
=
\dfrac{f_\vartheta '(\vartheta)}{\omega _0^2} \, .
\end{equation}

If $\vartheta \neq \pi/2$, Eq. (\ref{eq:pphicon}) can be solved for $p_{\varphi}/\omega _0$,
\begin{equation}
\dfrac{p_{\varphi}}{\omega _0} = 
\pm \, \mathrm{sin} \, \vartheta \,
\sqrt{a^2 \mathrm{sin} ^2 \vartheta + 
\dfrac{
f_{\vartheta} ' (\vartheta) \mathrm{sin} \, \vartheta
}{
2 \omega _0 ^2 \mathrm{cos} \, \vartheta
}}
\, .
\end{equation}
Inserting this expression into (\ref{eq:omega0con}) gives the region where conical light rays exist outside of the equatorial plane,
\[
(\rho ^2 - 2 m r ) 
\sqrt{a^2 \mathrm{sin} ^2 \vartheta + 
\dfrac{
f_{\vartheta} ' (\vartheta) \mathrm{sin} \, \vartheta
}{
2 \omega _0 ^2 \mathrm{cos} \, \vartheta
}}^{\, 2}
\]
\[
\pm \, 4 m r a \, \mathrm{sin} \, \vartheta 
\sqrt{a^2 \mathrm{sin} ^2 \vartheta + 
\dfrac{
f_{\vartheta} ' (\vartheta) \mathrm{sin} \, \vartheta
}{
2 \omega _0 ^2 \mathrm{cos} \, \vartheta
}}
\]
\begin{equation}
\le
(r^2 + a^2 ) \rho ^2
+ 2 m r a^2 \mathrm{sin} ^2 \vartheta
- \Delta \dfrac{f_r(r)+ f_{\vartheta} ( \vartheta )}{\omega _0^2} 
\end{equation}
We see that, as for the spherical light rays, there are at most two conical light rays through a point outside the equatorial plane.

If $\vartheta = \pi /2$, Eqs.(\ref{eq:Kcon}) and (\ref{eq:pphicon}) give the necessary conditions for light rays to exist in the equatorial plane. In other words, any equatorial ray is conical with $\vartheta = \pi /2$. In this case Eq.(\ref{eq:pphicon}) reduces to $f_{\vartheta} ' (\pi/2) =0$ which is equivalent to 
\begin{equation}\label{eq:plsymmetry}
\dfrac{\partial \omega _p(r, \vartheta) ^2}{\partial \vartheta} \Big| _{\vartheta = \pi/2} = 0 \, .
\end{equation}
If the plasma density does not satisfy this condition, a light ray that starts tangentially to the equatorial plane will not stay in this plane. In the following we assume that (\ref{eq:plsymmetry}) is satisfied and we write, for the sake of brevity, $\omega _p (r)$ instead of $\omega _p (r , \pi/2 )$, i.e.
\begin{equation} \label{eq:omega-f-r}
\omega _p (r) ^2  = \dfrac{f_r(r) + f_{\vartheta}(\pi /2)}{r^2} \, .
\end{equation}
Then for light rays in the equatorial plane the equations of motion (\ref{eq:dott}), (\ref{eq:dotphi}), (\ref{eq:dottheta}) and (\ref{eq:dotr}) reduce to
\begin{equation}\label{eq:dott2}
r^2 \,  \dot{t} =
\dfrac{  \left[  (r^2+a^2) r^2 + 2mra^2 \right] 
\omega _0 + 2mra p_{\varphi}}{c \, \Delta }
 \, ,
\end{equation}
\begin{equation}\label{eq:dotphi2}
r^2 \dot{\varphi}  =
\dfrac{2mra \, \omega _0
+ \big( r^2 - 2mr \big) p_{\varphi}}{
\Delta } \, , 
\end{equation}

\begin{equation}\label{eq:dottheta2}
0 = K 
-\left( p_{\varphi} -
a \, \omega _0 \right)^2
\, ,  
\end{equation}
\begin{equation}\label{eq:dotr2}
r^4 \dot{r}{}^2= - K \Delta +
 \left[ (r^2+a^2) \, \omega _0 - a p_{\varphi} \right] ^2
- \omega_p(r)^2 r^2 \, \Delta 
\, . 
\end{equation}
By solving (\ref{eq:dottheta2}) for the Carter constant $K$ and inserting the result into (\ref{eq:dotr2}) we find the equation for the $r$-coordinate in the form:
\[
r^4 \dot{r}{}^2  =   -  (p_\varphi - a \, \omega_0)^2  \Delta +
 \left[ (r^2+a^2) \, \omega _0 - a p_{\varphi} \right] ^2
\]
\begin{equation}\label{eq:dotr-new}
- \, \omega_p(r)^2 \, r^2 \, \Delta 
\, . 
\end{equation}

The system of equations (\ref{eq:dott2}), (\ref{eq:dotphi2}), (\ref{eq:dotr-new}) determines the motion of light rays in the equatorial plane of the Kerr metric surrounded by a plasma. Note that with $\omega_p(r) \neq 0$ the constants of motion $\omega_0$ and $p_\varphi$ enter separately whereas in the vacuum case the light rays are determined, up to parametrization, by the quotient $p_{\varphi}/\omega _0$. This reflects, of course, the fact that a plasma is a dispersive medium. 

Here we have derived Eq. (\ref{eq:dotr-new}) under the assumption that the plasma density in 3-dimensional space $\omega _p (r , \vartheta )$ satisfies the separability condition (\ref{eq:sepcon}). Actually, the light rays in the equatorial plane satisfy Eq. (\ref{eq:dotr-new}), with $\omega _p (r , \pi /2 )$ abbreviated as $\omega _p (r)$, even if this separability condition does not hold. It is only required that (\ref{eq:plsymmetry}) holds which guarantees that light rays remain in the equatorial plane if they start tangentially to it. In particular, this is true if the plasma density is symmetric with respect to the plane $\vartheta = \pi /2$, i.e. if $\omega _p (r , \vartheta) = \omega _p (r , \pi - \vartheta)$.


\section{Circular light rays on and off the equatorial plane}
\label{sec:circular}

In this Section we investigate the existence of light rays that are circular about the axis of symmetry. It is well known that in vacuum there are only two such circular light rays in the domain of outer communication of a Kerr black hole with $a \neq 0$. Both are in the equatorial plane and the inner one is co-rotating with the black hole while the outer one is counter-rotating. In the Schwarzschild case $a=0$ these two circular light rays are both located at $r=3m$ and, because of the spherical symmetry, any axis through the origin is a symmetry axis. (Beyond the horizon there are three more circular lightlike geodesics in a Kerr spacetime with $a \neq 0$, one in the equatorial plane and two off the equatorial plane, but they are of no relevance for observers outside of the outer horizon.)

In a plasma, the situation is different: circular light rays in the domain of outer communication may exist both in the equatorial plane and off the equatorial plane. We will now derive the necessary and sufficient conditions for the existence of circular light rays on and off the equatorial plane.

As a circular light ray lies at the intersection of a coordinate sphere $r = \mathrm{const.}$ and a coordinate cone $\vartheta = \mathrm{const.}$, it has to satisfy the two equations (\ref{eq:Ksph}) and (\ref{eq:pphispher}) for spherical light rays and the two equations (\ref{eq:Kcon}) and (\ref{eq:pphicon}) for conical light rays. If we eliminate the Carter constant $K$ from (\ref{eq:Ksph}) and (\ref{eq:Kcon}) and introduce the quantity
\begin{equation}
\xi = \dfrac{a p_{\varphi}}{\omega _0} - a^2 \mathrm{sin} ^2 \vartheta \, ,
\label{eq:defxi}
\end{equation}
we get a quadratic equation for $\xi$:
\begin{equation}
\xi ^2
=
\dfrac{
- a^2 \mathrm{sin}^2 \vartheta
}{
\rho ^2 - 2mr
}
\Big( 2 \rho ^2 \xi- r F( r , \vartheta ) \Big)
\, .
\label{eq:circpphi1}
\end{equation}
Here we have introduced the abbreviation
\begin{equation} 
F(r , \vartheta ) = \dfrac{1}{r} 
\left( 
\rho ^4 - \Delta \dfrac{f_r(r)+f_{\vartheta} ( \vartheta )}{\omega _0^2}
\right)
\end{equation}
and we have used that $\Delta - a^2 \mathrm{sin} ^2 \vartheta = \rho ^2 - 2mr$. We get a second quadratic equation for the same quantity $\xi$ from
(\ref{eq:pphispher}). Expressing everything with the same function $F(r , \vartheta )$, we find after some algebra
\[
\xi ^2 = 2 \xi \left( \rho ^2 - \dfrac{r \Delta}{r-m} \right)
- 
r F(r , \vartheta ) 
\]
\begin{equation}
+ \,
\dfrac{\Delta}{2(r-m)} 
\left( F(r, \vartheta ) + r \dfrac{\partial F(r, \vartheta )}{\partial r} \right) 
\, .
\label{eq:circpphi2}
\end{equation}
Subtracting (\ref{eq:circpphi2}) from (\ref{eq:circpphi1}) results in a linear 
equation for $\xi$,
\begin{equation}
\xi 
=
\dfrac{1}{4m} 
\left(
F(r,\vartheta ) - \dfrac{(\rho ^2-2mr) \, r}{(r^2-a^2 \mathrm{cos}^2 \vartheta )}
\dfrac{\partial F(r, \vartheta )}{\partial r}
\right)
\, .
\label{eq:xilin}
\end{equation}
Reinserting this expression for $\xi$ into (\ref{eq:circpphi1}) gives us an equation that does not contain $p_{\varphi}$ or $K$,
\[
8 \, m \, a^2 \mathrm{sin} ^2 \vartheta
\left(
\dfrac{\rho ^2 \, r}{(r^2-a^2 \mathrm{cos}^2 \vartheta )}
\dfrac{\partial F(r, \vartheta )}{\partial r}
-F(r,\vartheta )
\right)
\]
\begin{equation}
=
\left(
F(r,\vartheta ) - \dfrac{(\rho ^2-2mr) \, r}{(r^2-a^2 \mathrm{cos}^2 \vartheta )}
\dfrac{\partial F(r, \vartheta )}{\partial r}
\right) ^2
\, .
\label{eq:circgen}
\end{equation}
This is a necessary condition the coordinates $r$ and $\vartheta$ of a circular light ray have to fulfill, for given $\omega _0$. It is not sufficient because we also have to take Eq.(\ref{eq:pphicon}) into account. Here we have again to distinguish the two cases $\vartheta \neq \pi /2$ and $\vartheta = \pi /2$. In the first case (\ref{eq:pphicon}) can be divided by $\mathrm{cos} \, \vartheta$ which gives us another quadratic equation for $\xi$,
\begin{equation}
\xi ^2 
= 
-2 \, \xi \, a^2 \mathrm{sin} ^2 \vartheta 
+ 
\dfrac{
a^2 \mathrm{sin}^3 \vartheta \, f_{\vartheta} ' ( \vartheta )
}{
2 \, \omega _0^2 \mathrm{cos} \, \vartheta
}
\, .
\end{equation}
Subtracting this equation from (\ref{eq:circpphi1}) and inserting (\ref{eq:xilin}) 
results in 
\begin{equation}
\dfrac{r^2 \, \mathrm{cos} \, \vartheta}{(r^2-a^2 \mathrm{cos}^2 \vartheta )}
\dfrac{\partial F( r , \vartheta )}{\partial r}
=
\dfrac{
\mathrm{sin} \, \vartheta \, f_{\vartheta}' (\vartheta )
}{
2 \, \omega _0^2 
}
\, .
\label{eq:circoff}
\end{equation}
Equations (\ref{eq:circgen}) and (\ref{eq:circoff}) together are the necessary and sufficient conditions the coordinates $r$ and $\vartheta (\neq \pi /2)$ of a light ray off the equatorial plane have to satisfy. 

For $\vartheta = \pi /2$ we know already that (\ref{eq:pphicon}) holds if and only if $f_{\vartheta}' ( \pi /2 ) = 0$. If this condition is not satisfied, there are no light rays in the equatorial plane, in particular no circular ones. If it is satisfied, (\ref{eq:pphicon}) gives no further information.
Although (\ref{eq:circoff}) was derived under the assumption that $\vartheta \neq \pi /2$, it is actually true that (\ref{eq:circgen}) and (\ref{eq:circoff}) give the necessary and sufficient conditions for circular light rays in the equatorial plane as well:
By setting $\vartheta = \pi /2$ in (\ref{eq:circoff}) we see that the resulting equation is satisfied if and only if $f_{\vartheta}' ( \pi /2 ) = 0$, whereas (\ref{eq:circgen}) with $\vartheta = \pi /2$ simplifies to
\begin{equation}
8 m a^2 \Big(  r F_0'(r) -F_0(r) \Big)
=
\Big( F_0 (r) - (r-2m) F_0'(r) \Big) ^2
\label{eq:circ-final}
\end{equation}
where
\begin{equation} \label{eq:F-0-def}
F_0(r) = F(r , \pi /2 ) =
r^3 - r \Delta(r) \frac{\omega_p(r, \pi /2 )^2}{\omega_0^2} \, .
\end{equation}

We may summarize our findings in the following way: Equations (\ref{eq:circgen}) and (\ref{eq:circoff}) together are the necessary and sufficient conditions a circular light ray has to satisfy, both on and off the equatorial plane. For light rays on the equatorial plane these two equations simplify to equation (\ref{eq:circ-final}) and $f_{\vartheta} ' (\pi /2 ) = 0$.

With the coordinates $r$ and $\vartheta$ of a circular light ray known, and $\omega _0$ given, (\ref{eq:xilin}) with (\ref{eq:defxi}) determines the pertaining constant of motion $p_{\varphi}$, provided that $a \neq 0$, and (\ref{eq:Ksph}) or equivalently (\ref{eq:Kcon}) determines the pertaining constant of motion $K$. This is true for circular light rays on and off the equatorial plane. This demonstrates that for a Kerr black hole with $a \neq 0$ the constants of motion $p_{\varphi}$ and $K$ of a circular light ray at $(r, \vartheta )$ are unique. In the Schwarzschild case $a=0$ this is not trure, as we will see in Example 1 below.

Equation (\ref{eq:circ-final})
is equivalent to Eq.(43) of Paper I, if the latter is taken with the equality sign and simplified for the equatorial plane:
\[
\frac{a^2 r^2 \Delta }{(r-m)^2} \left( 1 \pm \sqrt{1 - f_r'(r) \frac{(r-m)}{2 r^2 \omega_0^2}  } \, \right)^2  - \dfrac{a^2  f_r(r) }{\omega _0^2} 
\]
\begin{equation} \label{eq:circ-Paper-I}
=
\left(\dfrac{r (a^2-m r)}{r-m}\pm \dfrac{r \Delta}{r-m}  
\sqrt{1 - f_r'(r) \frac{(r-m)}{2 r^2 \omega_0^2}  }\right) ^2 \, .
\end{equation}
In order to show the equivalence, it is necessary to write all terms in (\ref{eq:circ-Paper-I}) on a common denominator and then to take all terms that are proportional to plus-minus the square-root on one side. After squaring, the equation no longer contains radicals. It becomes possible to extract the factor $(r-m)^2$ in the numerator, which cancels with the same factor in the denominator. The resulting equation will be the same as Eq.(\ref{eq:circ-final}) if all terms are explicitly expanded.\\

\textit{Example 1}. Schwarzschild black hole. 

In the Schwarz\-schild spacetime ($a=0$) the separability condition (\ref{eq:sepcon}) simplifies to
\begin{equation}
    \omega _p (r , \vartheta )^2 = 
    \dfrac{f_r(r) + f_{\vartheta} (\vartheta)}{r^2} \, .
    \label{eq:Schwsep}
\end{equation}
For determining the circular light rays in this spacetime we observe that with $a=0$ Eq.(\ref{eq:circgen}) reduces to 
\begin{equation}
    r-3m = \dfrac{(r-2m)^2 f_r '(r)}{2 \, \omega _0 ^2 r^2} \, , 
\label{eq:Schwcirc1}
\end{equation}
and that, with this equation at hand, (\ref{eq:circoff}) can be rewritten as
\begin{equation}
\sin \vartheta \,
\dfrac{f_{\vartheta} ' (\vartheta )
}{
2 \, \omega _0^2}
= 
\mathrm{cos} \, \vartheta \, 
\Bigg( 
\dfrac{r^3}{(r-2m)} 
-
\dfrac{f_r(r)+ f _{\vartheta} ( \vartheta )}{\omega _0^2}
 \Bigg) \, .
 \label{eq:Schwcirc2}
\end{equation}
The pertaining constants of motion $p_{\varphi}$ and $K$ are then found by inserting this expression into (\ref{eq:pphicon}) and (\ref{eq:Kcon}),
\begin{equation} \label{eq:Schwpphi}
\dfrac{p_{\varphi}^2}{\omega _0^2}
=
\mathrm{sin} ^2 \vartheta
\left(
\dfrac{r^3}{r-2m} - \dfrac{f_r(r)+f_{\vartheta} ( \vartheta )}{\omega _0^2}
\right) \, ,
\end{equation}
\begin{equation}
\dfrac{K}{\omega _0^2}
=
\dfrac{r^3}{r-2m} - \dfrac{f_r(r)}{\omega _0^2}
\, .
\end{equation}
When deriving (\ref{eq:Schwpphi}) we have divided both sides of the equation by $\mathrm{tan} \, \vartheta$. However, by continuity the resulting equation is valid also for $\vartheta = \pi /2$. Note that (\ref{eq:Schwcirc1}) is independent of $\vartheta$. This equation determines the possible radius coordinate of circular light rays. If such a light ray actually exists at this radius value depends on whether (\ref{eq:Schwcirc2}), with this radius value, admits a solution $0<\vartheta < \pi$. In contrast to the Kerr case with $a \neq 0$, where every circular light ray comes with a unique $p_{\varphi}$ and a unique $K$, in the Schwarzschild case $p_{\varphi}$ is unique only up to sign, which demonstrates that a light ray may run through the circle either in positive or in negative $\varphi$ direction.

If $f_r$ and $f_{\vartheta}$ are identically zero, (\ref{eq:Schwcirc2}) requires $\mathrm{cos} \, \vartheta =0$,
i.e., in vacuum circular light rays exist only in the equatorial plane. However, a $\vartheta$-dependent plasma density can produce a kind of ``force'' on light rays that makes circular orbits about the symmetry axis off the equatorial plane possible.
More specifically, we read from (\ref{eq:Schwcirc2}) and (\ref{eq:Schwpphi}) that a circular light ray may exist above the equatorial plane if $f_{\vartheta }'$ is positive there and below the equatorial plane if $f_{\vartheta} '$ is negative there. 
E.g., for a plasma density $\omega _p (r , \vartheta ) ^2 / \omega _0^2 = 27 \, \sqrt{3}  \, m^2 \mathrm{sin} \, \vartheta \, \mathrm{cos} \, \vartheta / r^2$, which satisfies the separability condition with $f_r(r)=0$ and $f_{\vartheta} ( \vartheta ) / \omega _0^2= 27 \, \sqrt{3}  \, m^2 \mathrm{sin} \, \vartheta \, \mathrm{cos} \, \vartheta$, there is a circular light ray at $r=3m$ and $\vartheta = \pi /6$.

In the equatorial plane, circular light rays exist provided that $f_{\vartheta} ' ( \pi /2) = 0$ and their radius coordinate is then given by (\ref{eq:Schwcirc1}). At a colatitude coordinate $\vartheta \neq \pi /2$ the equation $f_{\vartheta} ' (\vartheta ) = 0$ can hold only for circular light rays with $p_{\varphi} = 0$, as can be seen from (\ref{eq:Schwcirc2}) and (\ref{eq:Schwpphi}). Such a light ray must have $\varphi = \mathrm{const.}$, as follows from (\ref{eq:dotphi}) with $a=0$, in addition to $r = \mathrm{const.}$ and $\vartheta = \mathrm{const.}$ which are the defining conditions for a circular light ray, so it does not really go around in a circle but is rather static. For such static light rays the right-hand side of Eq.(\ref{eq:Schwpphi}) vanishes. From Eq.(\ref{eq:n}) we see that the refractive index of the plasma at the location of such a light ray is equal to zero, $n=0$. This means, in particular, that the group velocity is zero: $v_{gr} = cn = 0$.
We may say that at points of the medium where $n>0$, a ray propagates with non-zero group velocity; at points of the medium where $n=0$, the ray stands (at least momentarily) still; points where $n<0$ cannot be reached by a ray. 
Static light rays, for which the gravitational attraction is exactly balanced by a ``force'' produced by the plasma, actually exist for some plasma densities. Here is an example. For a plasma density $\omega _p ( r , \vartheta )^2 /\omega _0 ^2 = 54 \, m^2 \mathrm{sin} \, \vartheta \, \mathrm{cos} \, \vartheta  / r^2$, which satisfies the separability condition with $f_r(r)=0$ and $f_{\vartheta} ( \vartheta ) / \omega _0 ^2= 54 \, m^2 \mathrm{sin} \, \vartheta \, \mathrm{cos} \, \vartheta$, there are static light rays at $r = 3m$ and $\vartheta = \pi /4$.

In the spherically symmetric case, i.e., if the separability condition (\ref{eq:Schwsep}) holds with $f_{\vartheta} (\vartheta )$ identically zero, (\ref{eq:Schwcirc2}) demonstrates that circular and non-static light rays exist only in the equatorial plane. As in this case any coordinate plane through the origin can be considered as the equatorial plane, this implies that all great circles (orthodromes) of the photon sphere are light rays. The radius $r_p$ of the photon sphere is given by (\ref{eq:Schwcirc1}). As an alternative, we can determine this radius by setting $a=0$ in Eq.(\ref{eq:circ-final}). This results in 
\begin{equation}
F_0(r) - F_0'(r) (r-2m) = 0 \, ,
\end{equation}
with the function $F_0(r)$ in the form
\begin{equation}
F_0(r) = r^3 - r^2 (r-2m) \frac{\omega_p(r)^2}{\omega_0^2} 
\end{equation}
which is indeed equivalent to (\ref{eq:Schwcirc1}). The case of a spherically symmetric plasma density on the Schwarzschild spacetime was already treated in our earlier paper \cite{Perlick-Tsupko-BK-2015} where we have written the equation of the photon sphere in the compact form
\begin{equation}
\frac{d}{dr} h(r)^2 = 0 \, ,
\end{equation}
where 
\begin{equation} \label{eq:function-h}
h(r)^2 = r^2 \left( \frac{r}{r-2m} - \frac{\omega_p(r)^2}{\omega_0^2}   \right) \, .
\end{equation}
Again, it is easy to check that this is equivalent to (\ref{eq:Schwcirc1}).

From Eq.(\ref{eq:Schwcirc1}), it follows that if $f_r'(r)=0$ (in particular, in the vacuum case when $f_r(r) \equiv 0$), then:
\begin{equation}
r_p = 3m \, .
\end{equation}
Correspondingly, for a plasma with plasma frequency
\begin{equation} \label{eq:power-law-2}
\omega_p (r)^2 = \frac{\mbox{Const}}{r^2}
\end{equation}
the radius of the photon sphere is exactly the same as in vacuum, regardless of the magnitude of the constant in Eq. (\ref{eq:power-law-2}).

In our paper \cite{Perlick-Tsupko-BK-2015} we have found the linear plasma corrections to the vacuum value of the photon sphere radius for power-law density distributions, $\omega_p^2 = C_0/ r^k$ with $k>0$. We have revealed that the linear correction is negative for $k>2$ and positive for $k<2$. In the borderline case, $k=2$, the linear correction is equal to zero. Interestingly, as shown above, in this specific case the radius is exactly equal to $3m$, even if we do not restrict to small densities.\\

\textit{Example 2}. Kerr black hole in vacuum.

We now briefly check that our equations reproduce the well known equations for circular vacuum light rays in the Kerr spacetime.  If $f_r(r)$ and $f_{\vartheta} ( \vartheta )$ are identically zero, Eq.(\ref{eq:circoff}) simplifies to 
\begin{equation}
\dfrac{
\rho ^2 \big( 3r^2-a^2 \mathrm{cos}^2 \vartheta \big) \mathrm{cos} \, \vartheta
}{
r^2-a^2 \mathrm{cos} ^2 \vartheta
}
=0 \, ,
\end{equation}
hence either $a^2 \mathrm{cos} ^2 \vartheta = 3r^2$ or $\vartheta = \pi /2$. In the first case substituting this expression for $a^2 \mathrm{cos} ^2 \vartheta$ into (\ref{eq:circgen}) results in
\begin{equation}
- \, 8ma^2 r^3 \mathrm{sin} ^2 \vartheta = 16 r^6 
\end{equation}
which can hold only for a negative value of $r$. This demonstrates that circular lightlike geodesics off the equatorial plane do not exist in the domain of outer communication of a Kerr black hole.

If $\vartheta = \pi /2$, we may use (\ref{eq:circ-final}). 
With $\omega_p(r) \equiv 0$, we write
\begin{equation}
F_0(r) = r^3 \, , \quad F_0'(r) = 3 r^2 \, ,
\end{equation}
and obtain
\begin{equation}
4 a^2 m   = r (r-3m)^2  
\end{equation}
or, in the form of Bardeen, Press and Teukolsky \cite{Bardeen-1972}, Eq.(2.17) there:
\begin{equation}
r^{3/2} - 3m r^{1/2} \pm 2 a m^{1/2}   = 0  \, .
\end{equation}
This cubic equation has two solutions in the domain of outer communication which give the well-known co-rotating and counter-rotating circular lightlike geodesics.\\

\textit{Example 3}. Kerr black hole and plasma with decreasing density with $k=2$ power-law index.

As another example we will now determine the circular light rays in the equatorial plane if the plasma density satisfies the separability condition with $f_r(r) =C_0$ and $f_{\vartheta} ( \vartheta )$ arbitrary except for the condition that $f_{\vartheta} ' ( \pi /2) = 0$ to make sure that light rays in the equatorial plane exist. We may then absorb the constant $f_{\vartheta} ( \pi /2 )$ into the constant $C_0$, i.e.,, we may assume that $\omega _p (r , \pi /2 ) ^2 = C_0 / r^2$. This case
is of particular interest, because then the plasma correction for a Schwarzschild black hole is zero, see above. Therefore, here a  plasma-correction to the circular photon orbit, which is at $3m$ in the Schwarzschild case, can occur only together with corrections due to the black-hole rotation. We will now calculate this correction for the case that the spin parameter $a$ is small.

We use $f_r'(r)=0$ in Eq.(\ref{eq:circ-Paper-I}). Also, as {demonstrated above, the plus sign has to be used for decreasing density profiles. We obtain:
\begin{equation} 
\frac{4 a^2 r^2 \Delta }{(r-m)^2}  - \dfrac{a^2  C_0 }{\omega _0^2}  =
\left(\dfrac{r^2(r-3m) + 2ra^2}{r-m} 
\right) ^2 \, ,
\end{equation}
which simplifies finally to
\begin{equation} 
r(r-3m)^2 - 4a^2 m + \dfrac{a^2 C_0 (r-m)^2}{\omega_0^2 r^3} = 0 \, .
\end{equation}
This is a sixth-order equation for $r$ which cannot be solved analytically. Therefore, we seek an approximate solution by linearizing with respect to $a$, i.e., we set
\begin{equation}
r = 3m + a \, r_1 \, , \quad a \ll m \, .
\end{equation}
We find that, to within this approximation, the counter-rotating and the co-rotating circular light rays in the equatorial plane are at
\begin{equation}
r_p = 3m \, \pm \, a \, \frac{2\sqrt{3}}{3} \sqrt{1 - \frac{C_0}{27m^2 \omega_0^2}} \, .
\end{equation}

\section{Comments on the black-hole shadow} 
\label{sec:shadow}

In this Section we return to the main topic of Paper I and give an additional discussion of the black-hole shadow. First, we remind the reader that for the construction of the shadow the photon region, as given in (\ref{eq:photonreg}), is crucial. In general, both signs in (\ref{eq:photonreg}) have to be taken into account. However, we have now proven that the minus sign can occur (i) only inside the ergoregion, (ii) only if $f_r'(r) <0$, and (iii) only if the plasma density is non-decreasing with respect to the radial coordinate, see the discussion below (\ref{eq:photonreg}). Moreover, in Paper I we have shown that the minus sign can be eliminated for a low-density plasma (see Section VI of Paper I) and for power-law distributions (see Eq.(79) in Section VIII of Paper I).

Here we want to use this opportunity for correcting an inaccuracy in Paper I. There we gave, in Eq.(72) the angular radius $\theta$ of the shadow of a Schwarzschild black hole in a plasma density $\omega _p (r, \vartheta )^2 = (f_r(r)+f_{\vartheta}(\vartheta ))/r^2$ as
\begin{equation}
\mathrm{sin} \, \theta= \sqrt{\dfrac{r_p^3 (r_O-2m)}{r_O^3 (r_p-2m)}}
\sqrt{
\dfrac{
1- \big( f_r(r_p)+f_{\vartheta} (\vartheta _O) \big) 
\frac{r_p-2m}{r_p^3 \omega _0^2}
}{
1- \big( f_r(r_O)+f_{\vartheta} (\vartheta _O) \big) 
\frac{r_O-2m}{r_O^3 \omega _0^2}
}
} \, .
\label{eq:SchwShadow}
\end{equation}
Here $r_p$ is the radius coordinate of the photon sphere and $r_O$ and $\vartheta _O$ are the coordinates of the observer position. From this equation, which is correct, we then concluded that the shadow is increased, in comparison to the vacuum case, if and only if 
\begin{equation}
\omega _p (r_p, \vartheta _O)^2 \Big( 1- \dfrac{2m}{r_p} \Big)  
< 
\omega _p (r_O, \vartheta _O)^2 \Big( 1- \dfrac{2m}{r_O} \Big)  
\, ,
\label{eq:ShadowCrit}
\end{equation}
i.e., if and only if the second square-root in (\ref{eq:SchwShadow} is bigger than 1. This conclusion is not quite correct because in (\ref{eq:SchwShadow}) the variable $r_p$ denotes the radius of the photon sphere in the plasma which, in general, is different from that in the vacuum, i.e., from $3m$. As the shadow radius $\theta _{\mathrm{vac}}$ in vacuum is given by
\begin{equation} \label{eq:shadow-vac}
\mathrm{sin} \, \theta _{\mathrm{vac}}   
=\sqrt{27 m^2 \, \dfrac{(r_O-2m)}{r_O^3}}
\, ,
\end{equation}
we have
\begin{equation}
\dfrac{
\mathrm{sin} \, \theta
}{
\mathrm{sin} \, \theta _{\mathrm{vac}}
}
=
\label{eq:ShadowRatio}
\end{equation}
\[
\sqrt{\dfrac{r_p^3}{27 m^2 (r_p-2m)}}
\sqrt{
\dfrac{
1- \big( f_r(r_p)+f_{\vartheta} (\vartheta _O) \big) 
\frac{r_p-2m}{r_p^3 \omega _0^2}
}{
1- \big( f_r(r_O)+f_{\vartheta} (\vartheta _O) \big) 
\frac{r_O-2m}{r_O^3 \omega _0^2}
}
}
\]
where the first square-root on the right-hand side is in general different from 1. However, if the plasma density is so low that we can linearize all equations with respect to $f_r$, $f_{\vartheta}$ and $f_r'$, we find from (\ref{eq:SchwSph}) that the radius of the photon sphere in the plasma can be approximated as
\begin{equation}
r_p = 3 m + \dfrac{f_r'(3m)}{18 \omega _0^2} \dots   
\, .
\end{equation}
This implies that, except for quadratic and higher-order terms, the first square-root in (\ref{eq:ShadowRatio}) equals 1, hence
\begin{equation}
\dfrac{
\mathrm{sin} \, \theta
}{
\mathrm{sin} \, \theta _{\mathrm{vac}}
}
=
\end{equation}
\[
1- 
\dfrac{
\omega _p (r_p, \vartheta _O) ^2 
}{ 
2 \, \omega _0^2
}
\Big( 1 - \dfrac{2m}{r_p} \Big)
+
\dfrac{
\omega _p (r_O, \vartheta _O) ^2 
}{ 
2 \, \omega _0^2
}
\Big( 1 - \dfrac{2m}{r_O} \Big)
\, \dots
\]
So in the linear approximation it is indeed true that the shadow is increased if and only if (\ref{eq:ShadowCrit}) holds. In the general case, however, one has to consider both square-roots in (\ref{eq:ShadowRatio}).

In Paper I we have observed that the boundary curve of the shadow may form ``fishtails'' if the plasma density is sufficiently big  (see Fig.6 there). We now add some comments on this situation. We reconsider the case of a constant plasma density, which is non-realistic in view of applications to astrophysics, but mathematically the simplest case for illustrating how the fishtails come about. As shown in Fig.$\,$5 of Paper I, in this case we have unstable spherical light rays (orange) between radii $r_{\mathrm{min}}$ and $r_{\mathrm{max}}$ and stable spherical light rays (hatched green) between radii $r_{\mathrm{s,min}}$ and $r_{\mathrm{s,max}}$. For $a = 0.999 m$ and $\omega _p^2 = 1.085 \, \omega _0^2$, the numbers are $r_{\mathrm{min}} \approx 1.0002  m$, $r_{\mathrm{max}} = r_{\mathrm{s,min}} \approx 8.4 m$ and $r_{\mathrm{s,max}} \approx 12  m$, see the middle panel of Fig.~5 in paper I. Note that the marginally stable spherical light ray at $r_{\mathrm{max}} = r_{\mathrm{s,min}}$ is \emph{not} circular, in contrast to the spherical light ray at $r_{\mathrm{min}}$.

For our fishtail shadow in Fig.$\,$6 of Paper I we chose the observer at $r_O = 5 m$ in the equatorial plane, i.e. the observer is inside the unstable photon region. In Figure \ref{fig:fishtails1} below we have decomposed the curve from Fig.$\,$6 in Paper I into several parts: Red for the part on one hemisphere, $0 <\theta <\pi /2$, and blue for the other hemisphere, $\pi /2 < \theta < \pi$. For either case, the boundary curve consists of three parts:

-- solid, where the constants of motion are the same as for an unstable spherical light ray at radius $r_p$ with $r_{\mathrm{min}} < r_p < r_O$;

-- dotted, where the constants of motion are the same as for an unstable spherical light ray at radius $r_p$ with $r_O < r_p < r_{\mathrm{max}} = r_{\mathrm{s,min}}$;

-- dashed, where the constants of motion are the same as for a stable spherical light ray at radius $r_p$ with $r_{\mathrm{max}} = r_{\mathrm{s,min}} < r_p < r_{\mathrm{s,max}}$.

\begin{figure}
	\includegraphics[width=0.85\columnwidth]{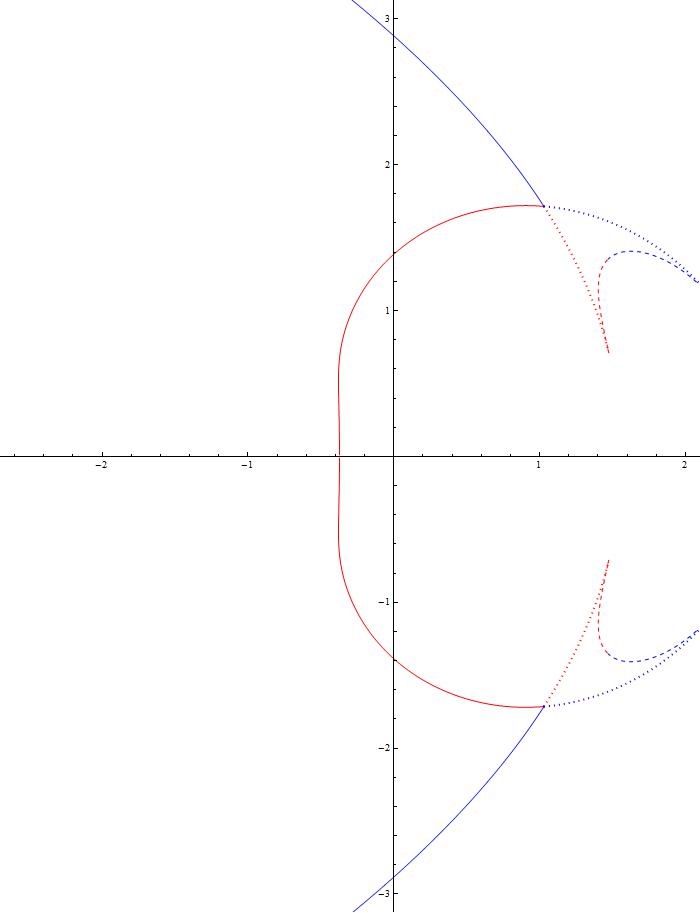}
    \caption{Shadow curve with fishtails, cf. Fig.6 in Paper I}
    \label{fig:fishtails1}
\end{figure}

Note the difference between the trapping mechanisms: Unstable spherical light rays are approached by other light rays from one side in an asymptotic spiral motion. Stable spherical light rays have trapped light rays in their neighborhood that oscillate about them.

The solid part of the boundary curve corresponds to light rays that approach an unstable spherical light ray from above, either directly (red) or after going through a maximum radius (blue). The dotted part corresponds to light rays that approach an unstable spherical light ray from below, either directly (blue) or after going through a minimum radius (red). The tips of the fishtails correspond to the light rays that approach the marginally stable light ray at $r_{\mathrm{max}} = r_{\mathrm{s,min}}$ from below.

In this situation we have two types of light rays issuing from the observer position into the past: Those which go to the horizon and those which oscillate between a minimum radius and a maximum radius forever. (For this choice of parameters there are no light rays going out to infinity.) For deciding about assigning darkness or brightness to a certain point on the observer's celestial sphere it is now of crucial relevance where the light sources are. There are several possibilities: 

(a) We may stick with the rule to associate darkness with all light rays that go to the horizon. Then we have to assume that there are no light sources in the region crossed by those light rays, in particular not in the region filled with stable spherical light rays. We are then forced  to associate darkness also with all light rays that oscillate between a minimum and a maximum. This would mean that for our observer placed inside the unstable photon region the entire sky is dark.

(b) We may assume that, among all light rays that go to the horizon, only those do not meet a light source that have either no turning point or a turning point below $r_{\mathrm{max}} = r_{\mathrm{s,min}}$. In agreement with this assumption we would have to assume that there are light sources in the region of stable spherical light rays and above, but not below. The resulting shadow is shown in Figure \ref{fig:fishtails2}. It looks  similar to the ordinary Kerr shadow in vacuo.

\begin{figure}
	\includegraphics[width=0.95\columnwidth]{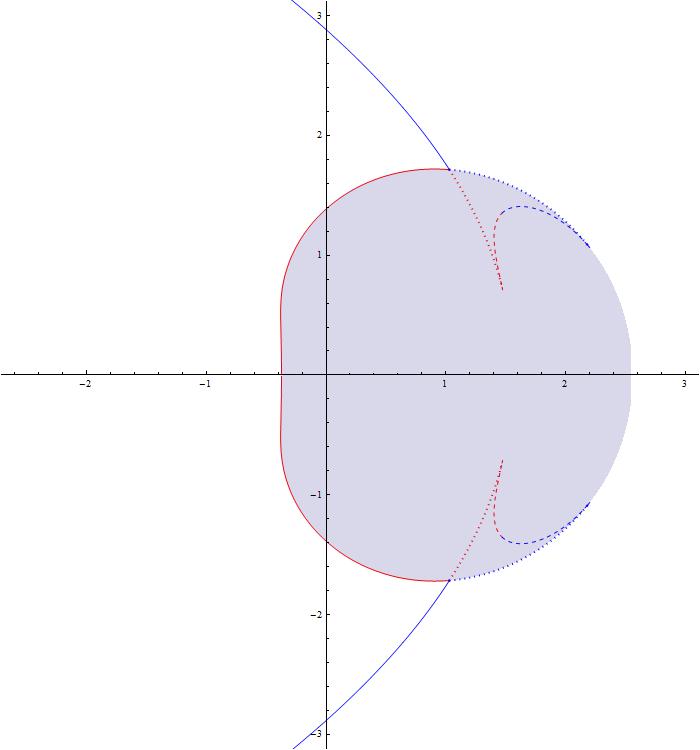}
    \caption{The shadow as it is actually seen, according 
    to assumption (b), if the fishtails are present.}
    \label{fig:fishtails2}
\end{figure}

The boundary of this shadow corresponds to light rays that spiral towards unstable spherical light rays (red solid and blue dotted curves), and of light rays that have their maximum radius exactly at $r_{\mathrm{max}} = r_{\mathrm{s,min}}$. Note that both on the upper half and on the lower half of the fishtail-curve there are \emph{two} points corresponding to the same radius value of the limiting spherical light ray. The shadow of Figure \ref{fig:fishtails2} singles out one of them: The point on the solid curve for $0 < \theta < \pi /2$ (red) and the point on the dotted curve for $\pi /2 < \theta < \pi$ (blue).

(c) We may assume that the region with light sources is not exactly bounded by  $r_{\mathrm{max}} = r_{\mathrm{s,min}}$ but rather by some other value $r_{\mathrm{limit}}$. As long as $r_{\mathrm{limit}}$ is bigger than $r_O$, the red solid curve is still part of the boundary of the shadow, as in Figure \ref{fig:fishtails2}, but the other part of the boundary is shifted either inwards or outwards. If $r_{\mathrm{limit}}$ is smaller than $r_O$, the entire sky is bright.

So the situation is rather complicated: The shadow very strongly depends on where the light sources are placed, and there is no general rule which assumption on the light sources is ``right'' and which is ``wrong''. As long as consistency is respected, mathematically one can choose any rule. However, whatever choice for the position of the light sources is  made, in \emph{no} case the observer will actually see the fishtails in the sky.

We emphasize that all six parts of the fishtail-curve (red/blue solid, red/blue dotted, red/blue dashed) correspond to light rays that have the same constants of motion $K$ and $p_{\varphi}$ as a spherical light ray. We now want to investigate which of the six parts correspond to light rays that actually approach a spherical light ray. To that end, we introduce for each $K$ and $p_{\varphi}$ the effective potential
\begin{equation}
V(r) =  K - \dfrac{1}{\Delta} \big( (r^2+a^2) \, \omega _0 +
a p_{\varphi} \big) ^2 + f_r (r) \, .
\end{equation}
Then Eq. (\ref{eq:dotr}), which is the same as Eq.(32) of Paper I up to the changed sign convention for $\omega _0$, may be rewritten in the form of an energy conservation law,
\begin{equation}
\dfrac{\rho ^4}{\Delta} \, \dot{r}{}^2 + V(r) = 0 \, .
\end{equation}
As the first term (the ``kinetic energy'') is non-negative, only the region where $V(r) \le 0$ is allowed. The boundary of this region gives a turning point if it is not an extremum and a spherical light ray otherwise; at a maximum, the spherical light ray is unstable and can be approached asymptotically by other light rays.

\begin{figure}
	\includegraphics[width=0.95\columnwidth]{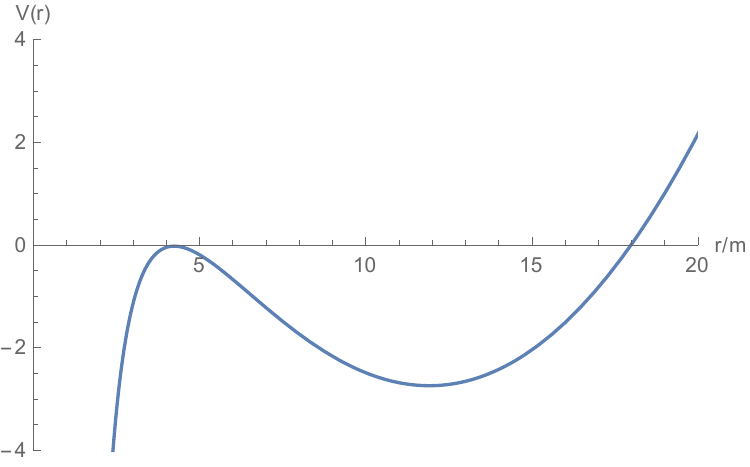}
    \caption{  Effective potential with a local maximum at a radius $r_p$ with $r_{\mathrm{min}} \approx 1.0002 m < r_p < r_O=5 m$.}
    \label{fig:Veff1}
\end{figure}

Figure \ref{fig:Veff1}  shows the potential for values of $K$ and $p_{\varphi}$ that correspond to an unstable spherical light ray at a radius $r_p$ with $r_{\mathrm{min}} < r_p < r_O$. The potential has a local maximum at $r_p$. We see that from the observer position at $r_O = 5 m$ there are two light rays with the chosen values of $K$ and $p_{\varphi}$: One that goes directly towards the unstable spherical light ray at $r_p$ (red solid curve in Figs. \ref{fig:fishtails1} and \ref{fig:fishtails2}) and another one that first moves outwards, has a turning point and then goes towards the unstable spherical light ray at $r_p$ (blue solid curve in Figs. \ref{fig:fishtails1} and \ref{fig:fishtails2}). If we perturb the potential a bit, so that the maximum goes either up or down, we see the following: The red solid curve separates light rays that go directly towards the horizon from light rays that oscillate between a maximum and a minimum. According to assumption (b) above we assign darkness to the first and brightness to the latter, so the red solid curve is part of the boundary curve of the shadow. Similarly, from the perturbed form of the potential we see that the blue solid curve separates light rays that have a turning point and then go to the horizon from light rays that oscillate between a maximum and a minimum. In any case, these neighboring light rays pass through the region of stable light rays. According to assumption (b) above we assign brightness to all of them, so the blue solid curve is not part of the boundary of the shadow. We repeat, however, that it corresponds to light rays that spiral towards an unstable spherical light ray.

Figure \ref{fig:Veff2} shows the effective potential for values of $K$ and $p_{\varphi}$ that correspond to an unstable spherical light ray at radius $r_p$ with $r_O < r_p < r_{\mathrm{max}}$. In this case, only the blue dotted curve corresponds to light rays that spiral towards the unstable spherical light ray at $r_p$. Points on the red dotted curve correspond to light rays with the same constants of motion; however, they go directly towards the horizon, so the red dotted curve is not part of the boundary curve of the shadow. Note that here it is essential that the potential has no zeros between the horizon and $r_p$.

\begin{figure}
	\includegraphics[width=0.95\columnwidth]{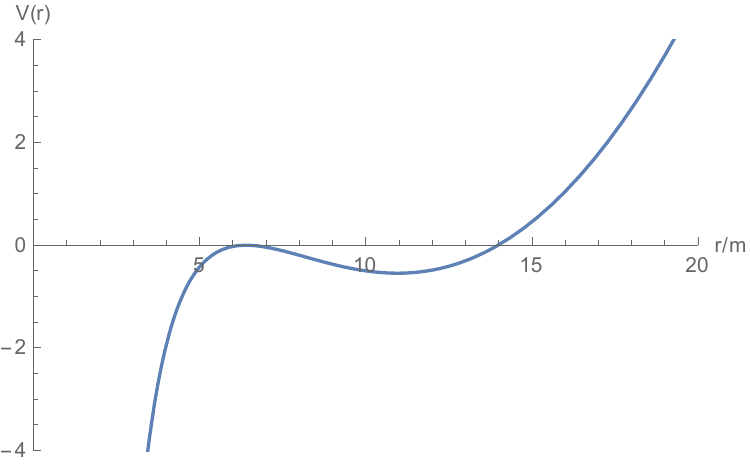}
    \caption{Effective potential with a local maximum at a radius $r_p$ with $r_O = 5 m < r_p < r_{\mathrm{max}} = r_{\mathrm{s,min}} \approx 8.4 m$.}
    \label{fig:Veff2}
\end{figure}

\begin{figure}
	\includegraphics[width=0.95\columnwidth]{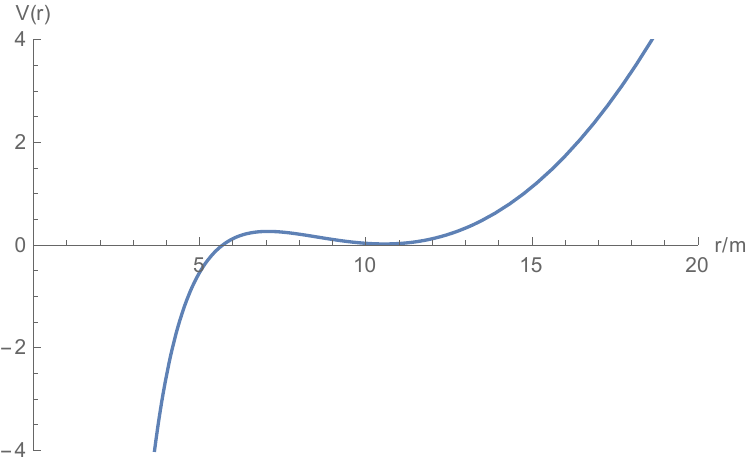}
    \caption{Effective potential with a local minimum at a radius $r_p$ with $r_{\mathrm{max}} = r_{\mathrm{s,min}}  \approx 8.4 m < r_p < r_{\mathrm{s,max}} \approx 12 m$.}
    \label{fig:Veff3}
\end{figure}

Figure \ref{fig:Veff3} shows the effective potential for values of $K$ and $p_{\varphi}$ that correspond to a stable spherical light ray at radius $r_p$ with $r_{\mathrm{s,min}}< r_p < r_{\mathrm{s,max}}$. In this case, the potential has a local minimum at $r_p$. Again, there are two light rays from the observer position with these constants of motion, corresponding to the red dashed and the blue dashed curve, respectively. The light ray on the red dashed curve goes directly towards the horizon, the light ray on the blue dashed curve has a turning point at a radius between $r_O$ and $r_{\mathrm{max}} = r_{\mathrm{s,min}}$. Neither of them comes close to $r_p$. According to assumption (b) we assign darkness to both of them.


\section{Exact deflection angle of light rays in the equatorial plane} \label{sec:exact}

\begin{widetext}

The equations of motion of a light ray in the equatorial plane of the Kerr metric in the presence of a non-homogeneous plasma are given by Eqs.(\ref{eq:dott2}), (\ref{eq:dotphi2}), (\ref{eq:dotr-new}), where it is assumed that (\ref{eq:plsymmetry}) holds and $\omega _p (r , \pi /2 )$ is abbreviated as $\omega _p (r)$. In this Section, we derive the expression for the deflection angle of such a light ray.

Dividing (\ref{eq:dotr-new}) by the square of (\ref{eq:dotphi2}) yields, after some elementary rearrangements, the orbit equation in the form

\begin{equation} \label{eq:orbitp-long} 
\left(
\dfrac{dr}{d \varphi} \right) ^2 = 
\dfrac{
\Delta (r)^2 
\left\{ 
r^3 - r \Delta (r) \dfrac{\omega _p (r)^2}{\omega _0^2} - \left( \frac{p_{\varphi}}{\omega_0} -a \right) \left[ 2 ar + (r-2m) \left( \frac{p_{\varphi}}{
\omega_0} -a \right) \right] 
\right\}
}
{ r 
\left[ ar +(r-2m) \left( \frac{p_{\varphi}}{\omega_0} -a \right) \right]^2
}
\, .  
\end{equation}
This equation agrees with Eq.(8.53) of Perlick \cite{Perlick-2000}. More generally, Crisnejos, Gallo and Jusufi \cite{Crisnejo-Gallo-Jusufi-2019} have presented the orbit equation of a light ray in the equatorial plane of a stationary and axisymmetric spacetime surrounded by a pressure-free non-magnetized plasma, see Eq.(106) there. If the Kerr metric coefficients (Eqs.(85), (86), (87) and (88) of their paper) are substituted, the resulting equation also agrees with our Eq.(\ref{eq:orbitp-long}.

As an alternative, the equation (\ref{eq:orbitp-long}) can be rewritten as
\begin{equation} \label{eq:orbitp-short}
    \left( \dfrac{dr}{d \varphi} \right) ^2 = \dfrac{\Delta (r)^2}{r (r-2m)} \left\{
    \dfrac{r^2 \Delta (r) \left[ 1 - \dfrac{\omega _p (r)^2}{\omega _0^2} \left(1-\dfrac{2m}{r} \right) \right]
    }{
    \left[ ar +(r-2m) \left( \frac{p_{\varphi}}{\omega_0} -a \right) \right]^2}
    -1 \right\} \, ,
\end{equation}
as can be easily verified by putting the two terms inside the bracket of (\ref{eq:orbitp-short}) on a common denominator and then pulling out a factor of $(r-2m)$ from the numerator.

Eq.(\ref{eq:orbitp-long}) is more involved than (\ref{eq:orbitp-short}), but it has the advantage of being directly applicable to light rays inside and outside of the ergoregion. Equations (\ref{eq:orbitp-short}) and (\ref{eq:orbitp-long}) are equivalent if $r \neq 2m$, but for light rays that cross the boundary of the ergoregion it is more convenient to use (\ref{eq:orbitp-long}) because it can be immediately applied whereas (\ref{eq:orbitp-short}) requires invoking the Bernoulli-l'H{\^o}pital rule or an analytic extension.

Either of the two equations (\ref{eq:orbitp-long}) and (\ref{eq:orbitp-short}) determines the geometric shape of all light rays in the equatorial plane that are not radial (i.e., for which $\varphi$ is not a constant). In particular, either of these equations gives us the deflection angle for any light ray in the equatorial plane that comes in from infinity, reaches a minimum radius value $r=R$ and then goes out to infinity again.  For such a light ray, the relation between the minimum radius $R$ and the constant of motion $p_{\varphi}/\omega _0$ can be found from the condition that the right-hand side of (\ref{eq:orbitp-short}) must vanish at $r=R$, i.e.
\begin{equation}
\frac{p_{\varphi}}{\omega_0} - a = w_{\pm} (R) 
\label{eq:pR}
\end{equation}
where 
\begin{equation} \label{eq:def-w}
w_{\pm} (R) = \frac{R}{R-2m}
\left\{ -a \pm \sqrt{\Delta (R)} \sqrt{ 1 - \dfrac{\omega _p(R)^2}{\omega _0^2} \Big(1- \dfrac{2m}{R} \Big) } \right\} \, .
\end{equation}
Here and in the following, the upper sign is for rays with $\dot{\varphi} / \omega _0 >0$ and the lower sign is for rays with $\dot{\varphi} / \omega _0 <0$, as can be read from (\ref{eq:dotphi2}). As we assume that $a \ge 0$, i.e., that the black hole is rotating in the positive $\varphi$ direction, this is equivalent to saying that the upper sign is for rays that are co-rotating and the lower sign is for rays that are counter-rotating with respect to the black hole. Inserting this expression for $p_{\varphi}/\omega _0$ into (\ref{eq:orbitp-short}) yields
\begin{equation}
    \left( \dfrac{dr}{d \varphi} \right) ^2 =
 \dfrac{ \Delta (r)^2 \left\{  r^3 - r \Delta (r) \dfrac{\omega _p (r)^2}{\omega _0^2} - w_{\pm} (R) 
 \Big( 2 ar + (r-2m) w_{\pm} (R) \Big) \right\}
}
{ r
\Big( ar +(r-2m) w_{\pm} (R) \Big)^2
}
  \, .
\label{eq:orbitp-long-with-R}  
\end{equation}
Solving for $d \varphi$ and integrating over the entire light ray gives us the angle $\Delta \varphi$ swept out by the light ray on its way from infinity to the minimum radius and back to infinity, 
\begin{equation} \label{angle-perlick}
\Delta \varphi \, = \, \pm \, 2 \int \limits_R^\infty
f(r) \, dr
\end{equation}
where
\begin{equation}
f(r) =  
\dfrac{ 
\sqrt{r} \,  \Big( ar +(r-2m) w_{\pm} (R) \Big)
}{ 
\Delta (r) \sqrt{  r^3 - r \Delta (r) \dfrac{\omega _p (r)^2}{\omega _0^2} - w_{\pm} (R) 
\Big( 2 ar + (r-2m) w_{\pm} (R) \Big) }
} \, .
\label{eq:deff}
\end{equation}
(This function $f(r)$ should not be confused with the function
$f_r(r)$ from Eq.(\ref{eq:sepcon}).)
$\Delta \varphi$ immediately gives us the deflection angle $\hat{\alpha}$ which is defined by
\begin{equation}
\hat{\alpha} = \pm \Delta \varphi - \pi  =  
2 \int \limits_R^\infty f(r) \, dr - \pi \, .
\label{eq:halpha}
\end{equation}
So $\hat{\alpha}$ is positive if the light ray is deflected towards the center, independently of whether the ray is co-rotating or counter-rotating.

Our new formula for $\Delta \varphi$ should be compared with an earlier formula that was given already in \cite{Perlick-2000}. After correcting a misprint (there was a spurious square in Eq. (8.55) that carried over into Eq. (8.57) of \cite{Perlick-2000}), this previous formula reads
\begin{equation} \label{eq:defl-perlick-2000}
\Delta \varphi = \pm \, 2 \int \limits_R^\infty \frac{\sqrt{r(r-2m)}}{r^2-2mr+a^2}   \left[  \frac{h(r)^2}{  \left( \frac{2ma}{r-2m} - \frac{2ma}{R-2m} \pm h(R) \right)^2 } -   1   \right]^{-1/2} dr \, ,
\end{equation}
where
\begin{equation}
h(r) = \left[  \frac{r(r^2-2mr+a^2)}{r-2m}  \left( \frac{r}{r-2m} - \frac{\omega_p(r)^2}{\omega_0^2} \right)  \right]^{1/2} \, .
\end{equation}

\end{widetext}
The main advantage of the new formula is in the fact that it can also be used for light rays that enter into the ergoregion, $R<r<2m$; in \cite{Perlick-2000} such light rays were excluded from the outset and the final formula (\ref{eq:defl-perlick-2000}) can be used inside the ergoregion only by analytic extension.

Below we will plot the exact deflection angle, for some special plasma densitites, both as a function of the minimum radius coordinate $R$ and as a function of the impact parameter $b$ which is often preferred for characterizing light rays that come in from infinity. To that end we need to establish the relation between the impact parameter and the constants of motion $p_{\varphi}$ and $\omega _0$.

In the vacuum case, for a light ray that comes in from infinity in an asymptotically flat spacetime, the impact parameter is known to be equal to $|p_\varphi/\omega_0 |$. To find out the physical meaning of $p_\varphi/\omega_0$ in the presence of a plasma, we consider the orbit equation in the form of Eq.(\ref{eq:orbitp-short}) in the situation that the light ray is still far away from the center. Assuming that the plasma density approaches a finite value for $r \to \infty$, (\ref{eq:orbitp-short}) can be written as:
\begin{equation} \label{impact-par-plasma-01}
\left( \dfrac{dr}{d \varphi} \right) ^2 = \frac{r^4}{(p_{\varphi}/\omega _0)^2} \left( 1 - \dfrac{\omega _p (\infty)^2}{\omega _0^2} \right) \Big( 1 + O \big( m/r \big) \Big)
\, .
\end{equation}
Integration yields
\begin{equation} \label{impact-par-plasma-02}
\underset{r \to \infty}{\mathrm{lim}} 
\Bigg(
 r^2 \Big(\varphi (r) - \varphi ( \infty ) \Big)^2 
 \Bigg)
= 
(p_{\varphi}/\omega _0)^2  \left( 1 - \dfrac{\omega _p (\infty)^2}{\omega _0^2} \right)^{-1} \, .
\end{equation}
The left-hand side is equal to the squared impact parameter $b^2$ of the light ray, so we find finally:
\begin{equation} \label{impact-par-plasma-03}
\left( \frac{p_\varphi}{\omega_0} \right)^2  = b^2 \left( 1 - \dfrac{\omega _p (\infty)^2}{\omega _0^2} \right) \, ,
\end{equation}
or
\begin{equation} \label{impact-par-plasma-04}
\left( \frac{p_\varphi}{\omega_0} \right)^2  = b^2 n(\infty)^2  \, .
\end{equation}
Formula (\ref{impact-par-plasma-04}) agrees with the result of \cite{Crisnejo-Gallo-Jusufi-2019}, see the text after their Eq.(106).
If $\omega_p(r) \to 0$ for $r \to \infty$, then the impact parameter has its usual relation with the constant of motion $(p_\varphi/\omega_0)$:
\begin{equation} \label{impact-par-plasma-05}
\left| \frac{p_\varphi}{\omega_0} \right|  = b  \, .
\end{equation}

We exemplify our new version (\ref{angle-perlick}) with (\ref{eq:deff}) of the deflection formula by Figures \ref{fig:Schwdefl} and \ref{fig:Kerrdefl} which show $\Delta \varphi$ in the Schwarzschild and in the extreme Kerr spacetime, respectively, for vacuum and for two different plasma densities.

From Figure \ref{fig:Schwdefl} we see that for large $R$ the presence of the plasma with a decreasing density profile makes the deflection smaller in comparison to the vacuum case. This can be easily explained by the refractive effect of a non-homogeneous plasma. We will further discuss this issue below for the case of weak deflection. At the same time, Figures \ref{fig:Schwdefl} and \ref{fig:Kerrdefl} reveal that for small enough $R$ the presence of the plasma can make the total deflection bigger in comparison with the vacuum case, even for a decreasing plasma density profile. Note that the deflection angle diverges if the light ray asymptotically approaches a circular light ray in the equatorial plane. If one takes this into account, one sees that for the Schwarzschild spacetime the above observation is in agreement with analytical results obtained earlier, see formulas (57) and (58) in \cite{Perlick-Tsupko-BK-2015}. In vacuum the radius of the photon sphere in the Schwarzschild metric equals $3m$. In \cite{Perlick-Tsupko-BK-2015} we have found analytically the linear correction to this radius due to the presence of a small-density plasma. It has been shown that, depending on the index $k$ of a power-law density profile, the radius of the photon sphere may become smaller ($k>2$) or bigger ($k<2$) than $3m$. This agrees with the behavior of the curves in Figure \ref{fig:Schwdefl} for small $R$.

\begin{figure}
	\includegraphics[width=\columnwidth]{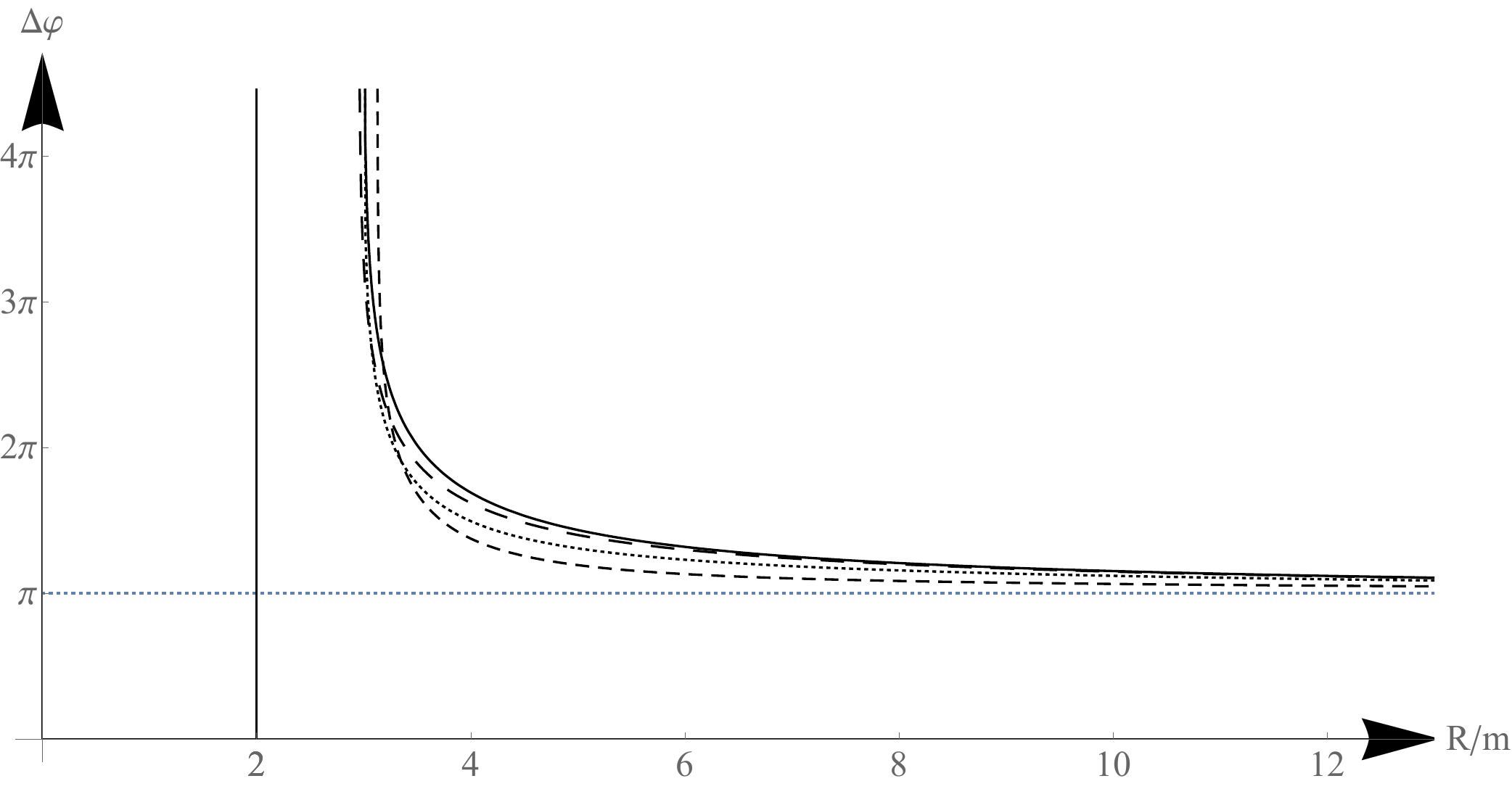}
    \caption{$\Delta \varphi$ in the Schwarzschild spacetime as a function of the minimum radius coordinate $R$, for light rays in vacuum (solid) and in a plasma with $\omega _p (r)^2 = 7 \, \omega _0 ^2 (m/r)^{3}$ (wide-dashed), $\omega _p (r)^2 = 7 \, \omega _0 ^2 (m/r)^{2}$ (dotted) and $\omega _p (r)^2 = 7 \, \omega _0 ^2 (m/r)^{3/2}$ (narrow-dashed). The vertical line marks the horizon.   }
    \label{fig:Schwdefl}
\end{figure}

\begin{figure}
	\includegraphics[width=\columnwidth]{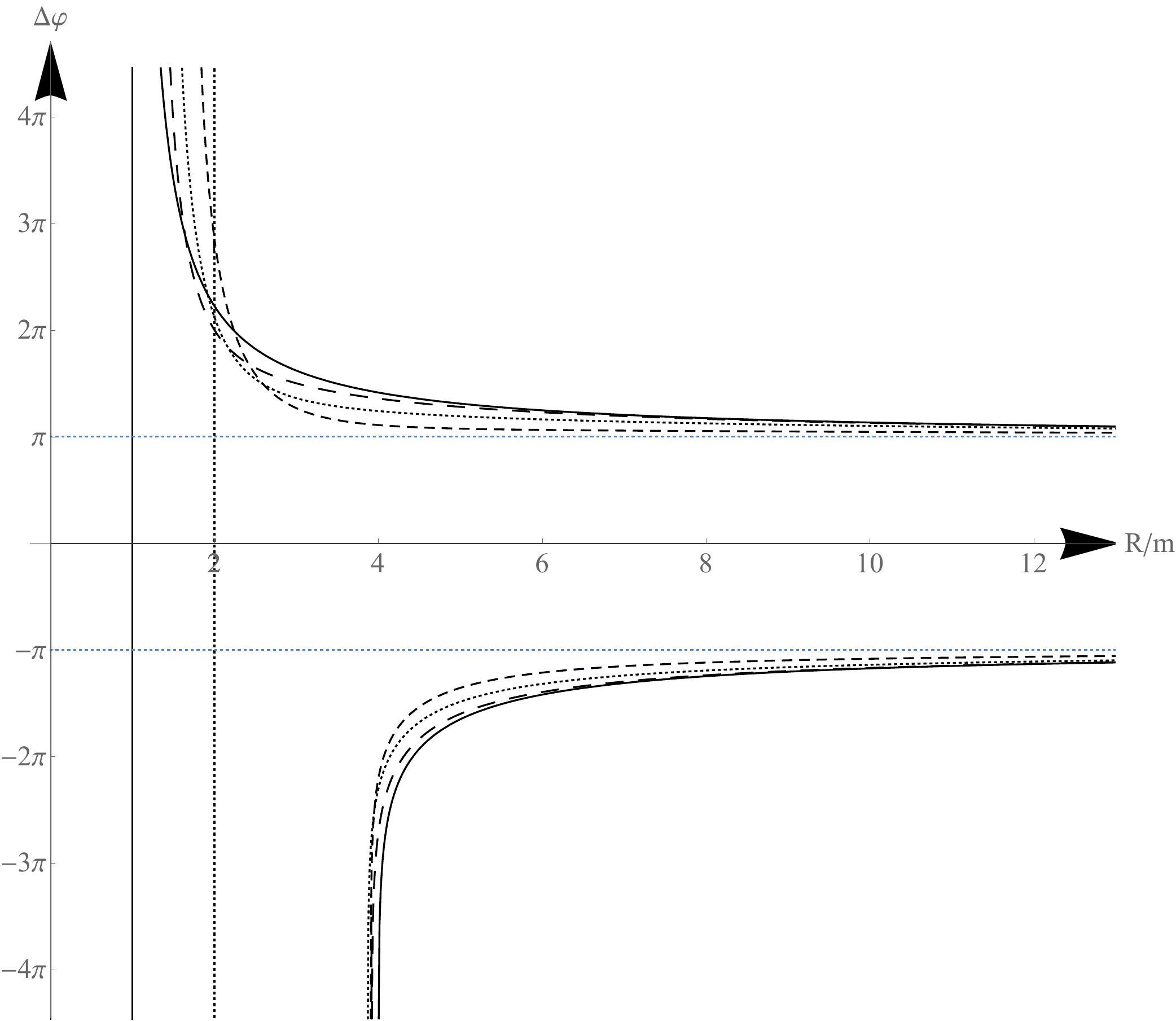}
    \caption{$\Delta \varphi$ in the extreme Kerr spacetime, $a=m$, as a function of the minimum radius coordinate $R$, for light rays in vacuum (solid) and in a plasma with $\omega _p (r)^2 = 7 \, \omega _0 ^2 (m/r)^{3}$ (wide-dashed), $\omega _p (r)^2 = 7 \, \omega _0 ^2 (m/r)^{2}$ (dotted) and $\omega _p (r)^2 = 7 \, \omega _0 ^2 (m/r)^{3/2}$ (narrow-dashed). The vertical solid line marks the (outer) horizon and the vertical dotted line marks the boundary of the ergoregion.    }
    \label{fig:Kerrdefl}
\end{figure}

With the help of (\ref{eq:pR}) the angle $\Delta \varphi$ (and, thereby, the deflection angle) can also be viewed as a function of the constant of motion $\big| p _{\varphi}  / \omega _0 \big|$, rather than as a function of the minimum radius $R$, see Figures \ref{fig:Schwdef2} and \ref{fig:Kerrdef2}. We have already demonstrated that in the case that $\omega _p (r)$ falls off to zero for $r \to \infty$ this constant of motion equals the impact parameter $b$. In Figures \ref{fig:Schwdef2} and \ref{fig:Kerrdef2} we see an important difference in comparison to the case where the minimum radius $R$ was plotted on the horizontal axis: Now there is no crossing of the graphs for different density profiles.

Qualitatively new features arise if the plasma is very dense or, what amounts to the same, if $\omega _0$ is very small. We illustrate this in Figs. \ref{fig:Schwdef3} and \ref{fig:Schwdef4} with another example on the Schwarzschild spacetime. We see that the wide-dashed plot shows a non-monotonic dependence of $\Delta \varphi$ on $R$ and also on $b$. Moreover, $\Delta \varphi$ is smaller than $\pi$ for some values of $R$ and $b$, i.e., it shows a negative deflection angle which means that the light ray is not attracted towards the center but rather repelled. In the narrow-dashed plot in Figs. \ref{fig:Schwdef3} and \ref{fig:Schwdef4}, which refers to an even denser plasma, $\Delta \varphi$ even goes to zero for $b \to 0$ which means that the light ray which aims directly at the center is reflected back in the same direction from which it came. At the point of reflection the index of refraction is zero and the light ray stands momentarily still. This phenomenon is well known e.g. from Earth's ionosphere where rays of very low frequency are reflected back.

\begin{figure}
	\includegraphics[width=\columnwidth]{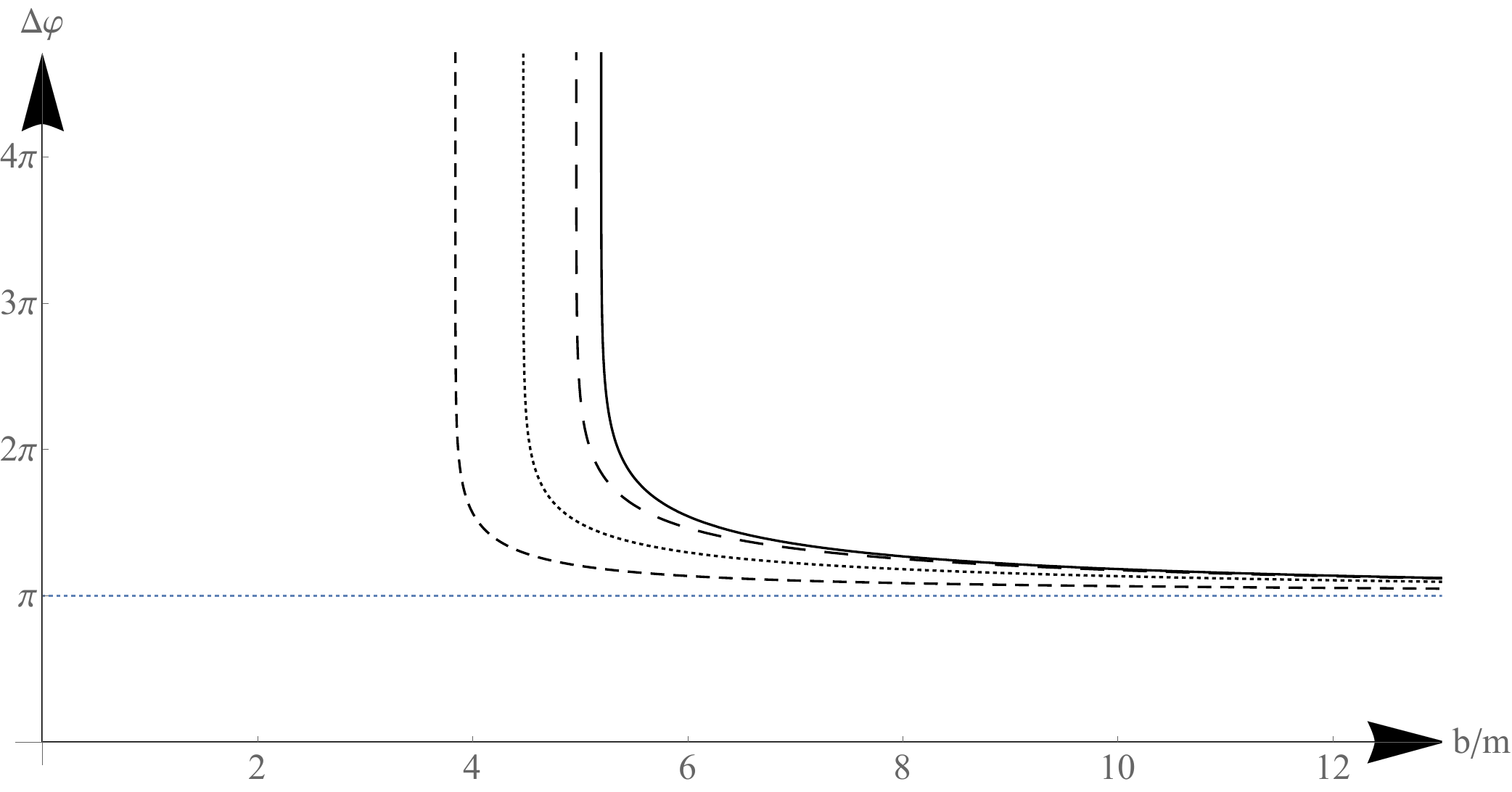}
    \caption{$\Delta \varphi$ in the Schwarzschild spacetime as a function of the impact parameter $b$, for light rays in vacuum (solid) and in a plasma with $\omega _p (r)^2 = 7 \, \omega _0 ^2 (m/r)^{3}$ (wide-dashed), $\omega _p (r)^2 = 7 \, \omega _0 ^2 (m/r)^{2}$ (dotted) and $\omega _p (r)^2 = 7 \, \omega _0 ^2 (m/r)^{3/2}$ (narrow-dashed).}
    \label{fig:Schwdef2}
\end{figure}

\begin{figure}
	\includegraphics[width=\columnwidth]{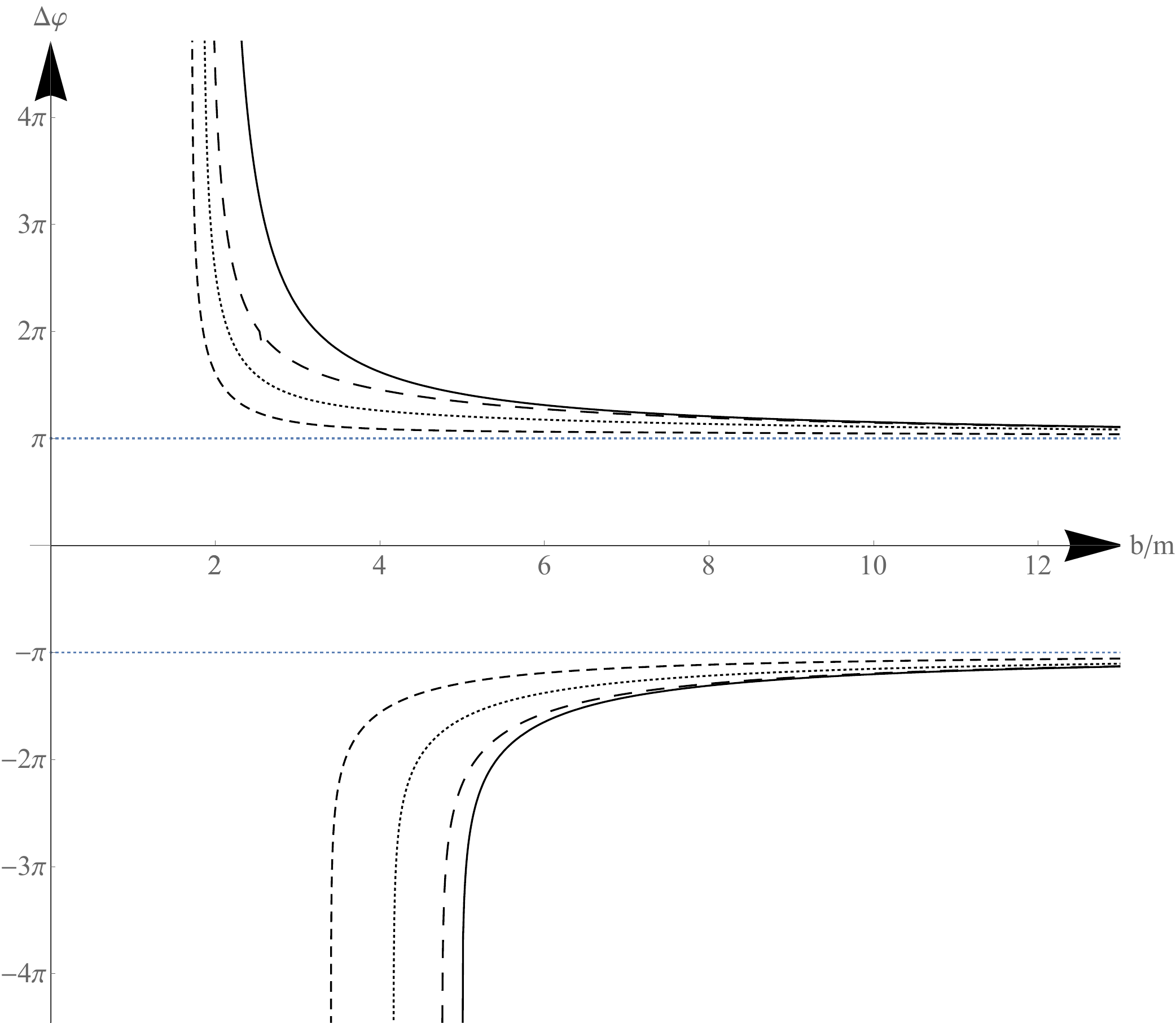}
    \caption{$\Delta \varphi$ in the extreme Kerr spacetime, $a=m$, as a function of the modulus of the impact parameter $|b|$, for light rays in vacuum (solid) and in a plasma with $\omega _p (r)^2 = 7 \, \omega _0 ^2 (m/r)^{3}$ (wide-dashed), $\omega _p (r)^2 = 7 \, \omega _0 ^2 (m/r)^{2}$ (dotted) and $\omega _p (r)^2 = 7 \, \omega _0 ^2 (m/r)^{3/2}$ (narrow-dashed). }
    \label{fig:Kerrdef2}
\end{figure}

\begin{figure}
	\includegraphics[width=\columnwidth]{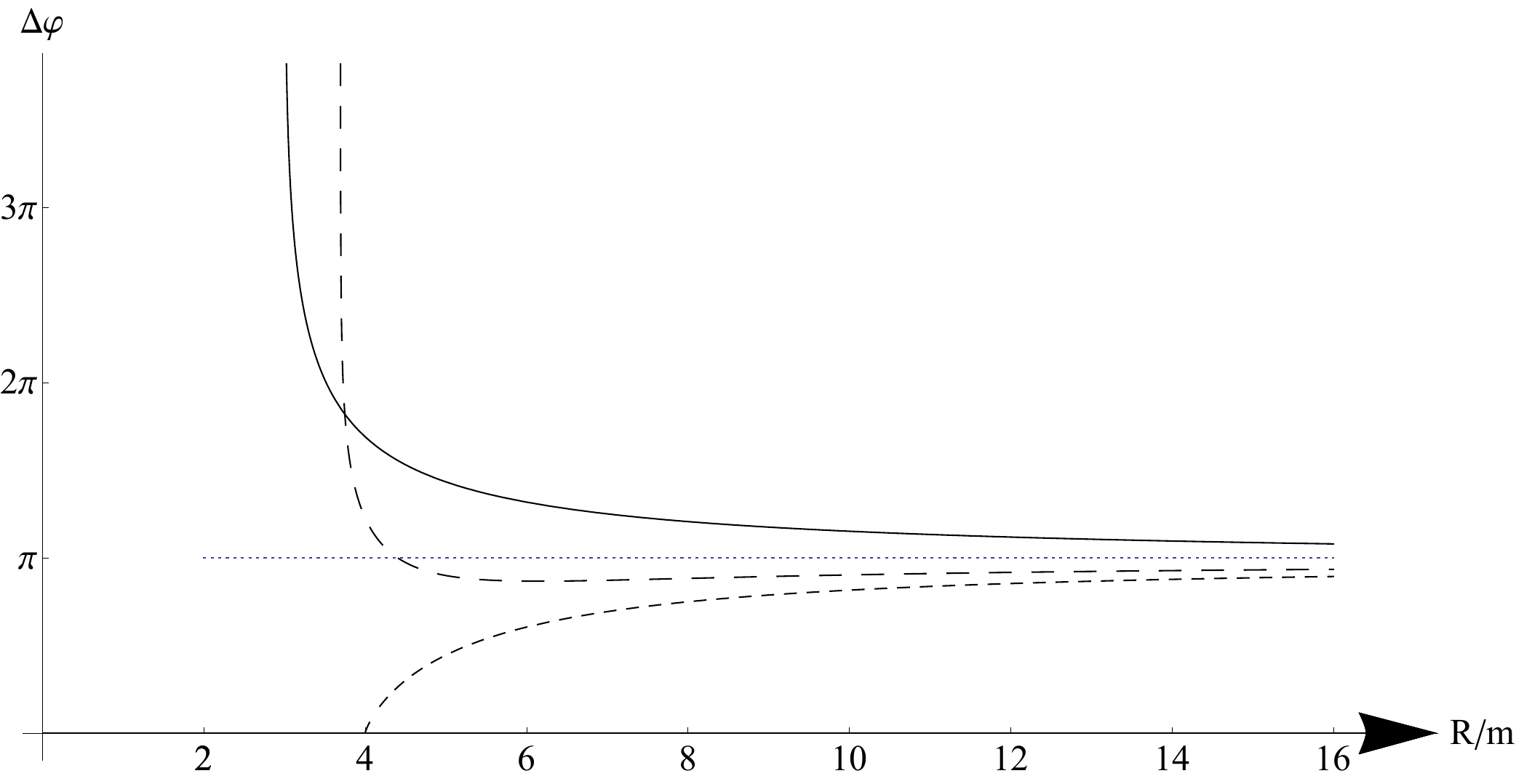}
    \caption{$\Delta \varphi$ in the Schwarzschild spacetime, $a=0$, as a function of the minimum radius $R$, for light rays in vacuum (solid) and in a plasma with $\omega _p (r)^2 = 6.6 \, \omega _0 ^2 m/r$ (wide-dashed), $\omega _p (r)^2 = 8 \, \omega _0 ^2 m/r$ (narrow-dashed). }
    \label{fig:Schwdef3}
\end{figure}

\begin{figure}
	\includegraphics[width=\columnwidth]{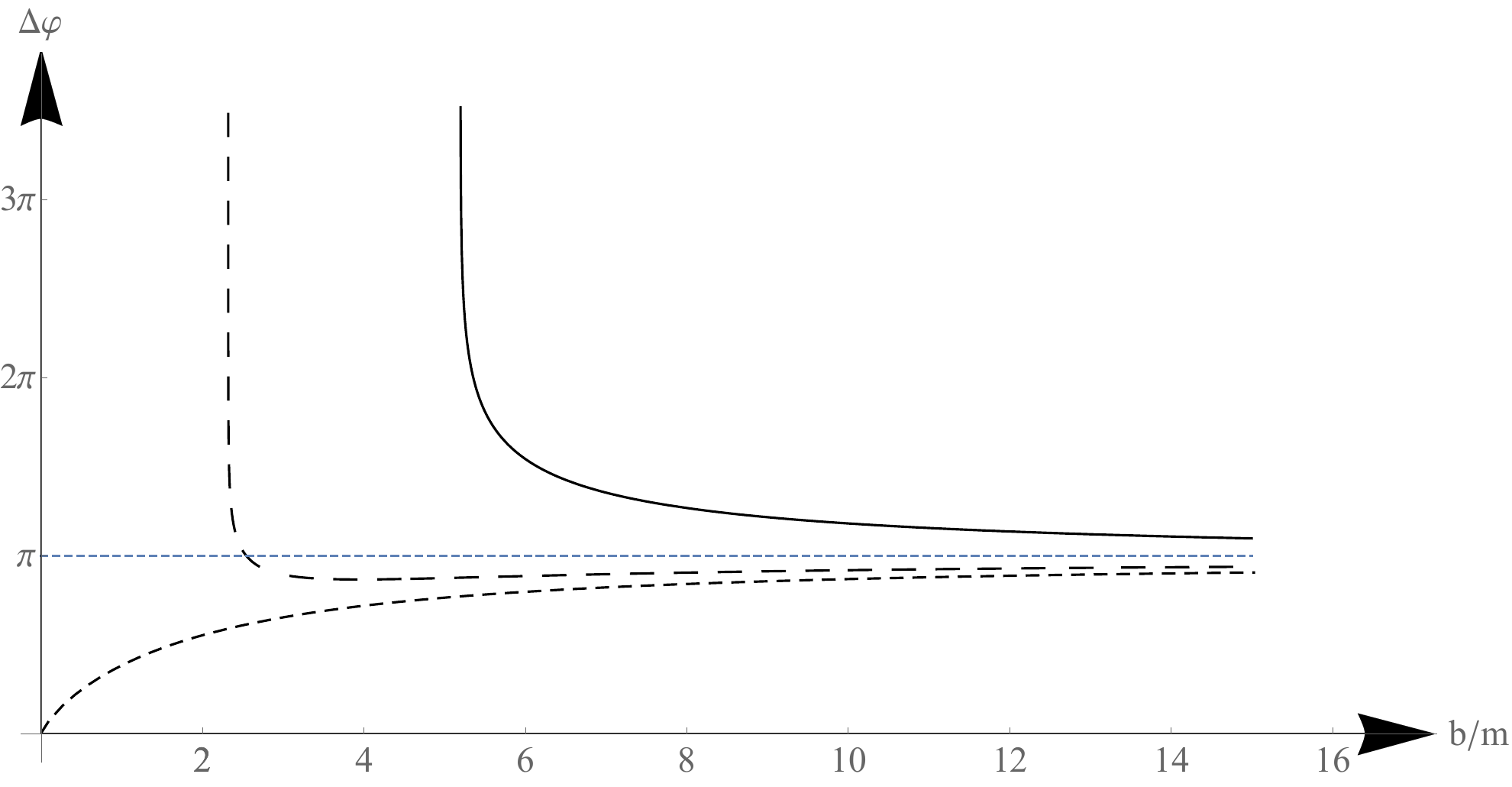}
    \caption{$\Delta \varphi$ in the Schwarzschild spacetime, $a=0$, as a function of the impact parameter $b$, for light rays in vacuum (solid) and in a plasma with $\omega _p (r)^2 = 6.6 \, \omega _0 ^2 m/r$ (wide-dashed), $\omega _p (r)^2 = 8 \, \omega _0 ^2 m/r$ (narrow-dashed). }
    \label{fig:Schwdef4}
\end{figure}

Without a plasma ($\omega _p (r) = 0$) Eq.(\ref{eq:halpha}) reduces to 
\begin{equation}
\hat{\alpha}{}_{\mathrm{vac}} = \pm  \Delta \varphi _{\mathrm{vac}}  - \pi 
\end{equation}
where 
\begin{equation}
\Delta \varphi _{\mathrm{vac}} = \pm 2 \,  \int \limits_R^\infty
f_{\mathrm{vac}}(r) \, dr  
\end{equation}
with
\begin{equation}
f_{\mathrm{vac}}(r) = \dfrac{\sqrt{r(r-2m)}}{r^2-2mr+a^2} \times 
\end{equation}
\[
\Bigg( 
\dfrac{(R-2m)^2r^2(r^2-2mr+a^2)}{(r-2m)R\sqrt{R^2-2mR+a^2} \mp \big(2ma(r-R) \big)^2}
\, -\, 1 \, \Bigg)^{-1/2} \, .
\]
Then
\begin{equation} \label{alpha-pl-definition}
\hat{\alpha}{}_{\mathrm{pl}} = \hat{\alpha} - 
\hat{\alpha}{}_{\mathrm{vac}} = \pm \Delta \varphi  \mp \Delta \varphi {}_{\mathrm{vac}}
\end{equation}
gives the contribution to the deflection angle that is produced by the effect of the plasma. If the frequency $\omega _0$ is big in comparison with the plasma frequency $\omega _p (r)$ everywhere along the ray, we may be satisfied with linearizing with respect to $\omega _p (r) ^2 / \omega _0^2 $ for all $R < r < \infty$. This results in \begin{equation} \label{alpha-pl-explicit}
\hat{\alpha}{}_{\mathrm{pl}} = 2 \, \int \limits_R^\infty
f_{\mathrm{pl}}(r) \, dr \, + \, \dots   
\end{equation}
where
\begin{equation}
f_{\mathrm{pl}}(r) = \dfrac{-f_{\mathrm{vac}} (r)^3 \Delta (r) ^3 (R-2n)^2r}{2m
\Big[(r-2m)R\sqrt{\Delta (R)} \mp 2ma(r-R) \Big]^3} \times
\end{equation}
\[
\Bigg\{\Big[(r-2m)R\sqrt{\Delta (R)} \mp 2ma(r-R) \Big] \dfrac{\omega_p(r)^2}{\omega _0^2} -
\]
\[
\sqrt{\Delta (R)} (R-2m) R \, \dfrac{\omega _p (R)^2}{\omega _0^2}
\Bigg\} \, .
\]


\section{Weak deflection of light rays in the equatorial plane}
\label{sec:weak}

In this Section we obtain the weak deflection approximation for the deflection angle $\hat{\alpha}$ introduced in (\ref{eq:halpha}).

In the situations considered above, the effects from the black hole spin and from the plasma environment of the black hole are mixed together. In order to separate the contributions of these two effects and to compare them with each other, we consider light rays flying past the black hole with big impact parameters which results in weak gravity effects on the ray. Additionally we assume that the plasma effects are also small in these regions which is true if the plasma density falls off if infinity is approached. This causes the deflection angle to be small, and the contributions to the deflection angle from gravity and from the plasma can then be separated from each other.

For calculation of the deflection angle in the case of weak deflection we will use the formula (\ref{eq:defl-perlick-2000}). We refer to the integrand of Eq.(\ref{eq:defl-perlick-2000}) as to $f(r)$. For further convenience, we also introduce the abbreviation $\beta_p(r) = \omega_p^2(r) / \omega_0^2$. We assume that the values of the dimensionless quantities $m/r$, $a/r$ and $\beta_p(r)$ are small. Then, the trajectory of a light ray is almost a straight line, i.e., $\hat{\alpha} \ll 1$.

We expand the expression $f(r)$ keeping terms proportional to $m$, $m^2$, $am$, $a^2$ and also $\beta_p(r)$ and $\beta_p(R)$. We obtain:
\begin{equation}
f(r) = f_0(r) + f_m(r) + f_{m2} (r) \, +
\end{equation}
\[
+ \, f_a(r) + f_{a2}(r) + f_\omega(r) \, ,
\]
where
\begin{equation}
f_0(r) = \frac{R}{r\sqrt{r^2-R^2}} \, ,
\end{equation}
\begin{equation}
f_m(r) = \frac{m (R^2+rR+r^2)}{r^2 (r+R) \sqrt{r^2-R^2} }   \,  ,
\end{equation}
\begin{equation}
f_{m2}(r) = \frac{3m^2 (R^2+rR+r^2)^2}{2r^3 R (r+R)^2 \sqrt{r^2-R^2} }   \,  ,
\end{equation}
\begin{equation}
f_a(r) = \mp \frac{2am}{R(r+R) \sqrt{r^2-R^2}}   \,  ,
\end{equation}
\begin{equation} \label{eq:f-a2-integrand}
f_{a2}(r) = \frac{a^2(r^2-2R^2)}{2R r^3 \sqrt{r^2-R^2}}  \,  ,
\end{equation}
\begin{equation}
f_\omega(r) = \frac{rR[\beta_p(r)-\beta_p(R)]}{2(r^2-R^2)^{3/2}}  \,  .
\end{equation}

By integration of these terms (see Eq.(\ref{eq:defl-perlick-2000})), we find the contributions of different effects to the total light deflection:
\begin{equation}
2 \int \limits_R^\infty f_0(r) \, dr = \pi \, ,
\end{equation}
\begin{equation}
2 \int \limits_R^\infty f_m(r) \, dr = \frac{4m}{R} \, ,
\end{equation}
\begin{equation}
2 \int \limits_R^\infty f_{m2}(r) \, dr = \frac{m^2(15 \pi - 16)}{4R^2} \, ,
\end{equation}

\begin{equation}
2 \int \limits_R^\infty f_{a}(r) \, dr = \mp \frac{4am}{R^2} \, ,
\end{equation}
\begin{equation} \label{f-a2-integral}
2 \int \limits_R^\infty f_{a2}(r) \, dr = 0 \, .
\end{equation}
Note that, although the terms proportional to $a^2$ are present in the integrand (see Eq.(\ref{eq:f-a2-integrand})), they do not contribute to the total deflection (see Eq.(\ref{f-a2-integral})). It is known that in the vacuum deflection formula of a Kerr black hole, terms with $a^2$ appear only in terms which are of third order altogether, as $ma^2/R^3$ or $ma^2/b^3$, see, e.g., Refs.\cite{Iyer-Hansen-2009a, Aazami-Keeton-2011b, Congdon-2022-kerr-bending}. Integration of the term $f_\omega(r)$ gives the refractive deflection of a photon:
\begin{equation}
\hat{\alpha}_{\mathrm{refr}} = 2 \int \limits_R^\infty f_{\omega}(r) \, dr \, .
\end{equation}
This expression can be transformed (see Appendix B in \cite{BK-Tsupko-2015}) into:
\begin{equation} \label{alpha-refr-R}
\hat{\alpha}_{\mathrm{refr}}(R) =  \frac{R K_e}{\omega_0^2} \int \limits_R^\infty \frac{1}{\sqrt{r^2-R^2}} \frac{dN(r)}{dr} \, dr \, , \; K_e \equiv \frac{4 \pi e^2}{m_e}
\end{equation}
where $N(r)$ is the number density of electrons in the equatorial plane.
As a result, we obtain the deflection angle as a function of $R$:
\begin{equation} \label{total-deflection-R}
\hat{\alpha}(R) = \frac{4m}{R} + \left( \frac{15 \pi}{4} - 4 \right) \frac{m^2}{R^2} \, \mp \,  \frac{4ma}{R^2} \, + \, \hat{\alpha}_{\mathrm{refr}}(R)  \,  .
\end{equation}
Here the first three terms describe the vacuum gravitational deflection. More precisely, the first term is the first-order Schwarzschild deflection, sometimes called the Einstein angle, and the second term is the second-order contribution of the Schwarzschild deflection. It makes sense to take this second-order term into account, as the rotation term is of the second order as well. The third term describes the deflection due to the rotation of the black hole, cf., e.g., \cite{Plebanski-1960, Cohen-Brill-1968}. In the presence of rotation, the deflection angle becomes smaller for direct orbits (co-rotation, upper sign), and becomes larger for retrograde motion (counter-rotation, lower sign), cf. 
\cite{Plebanski-1960, Cohen-Brill-1968, Cohn-1977, Dymnikova-1986, Edery-2006-kerr, Sereno-2006-kerr-weak}. 
(Note that some authors incorporate the plus-minus sign into the spin parameter $a$ with the convention that positive $a$ corresponds to co-rotation and negative $a$ corresponds to counter-rotation. However, we do not adopt this convention; as indicated above, we choose $a \ge 0$ throughout.)
The last term in (\ref{total-deflection-R}) is the refractive deflection, $\hat{\alpha}_{\mathrm{refr}}(R)$, given by Eq. (\ref{alpha-refr-R}).

For the gravitational lens equation, it is necessary to have the deflection angle not as a function of the distance of closest approach $R$ but of the impact parameter $b$. Up to quadratic terms, the gravitational deflection angle in vacuum is (see, for example, \cite{Edery-2006-kerr, Iyer-Hansen-2009a, Aazami-Keeton-2011b}):
\begin{equation}
\hat{\alpha}(b) = \frac{4m}{b} + \frac{15 \pi}{4} \frac{m^2}{b^2} \mp \frac{4ma}{b^2} .
\end{equation}
In turn, the refractive deflection $\hat{\alpha}_{\mathrm{refr}}$ can also be rewritten as a function of $b$ (see Appendix B in \cite{BK-Tsupko-2015} for details of the transformation):
\begin{equation} \label{alpha-refr-b}
\hat{\alpha}_{\mathrm{refr}}(b) =  \frac{K_e}{\omega_0^2}   \int \limits_0^\infty \frac{\partial N}{\partial b} \, dz \, .
\end{equation}
To link up with the notation of \cite{BK-Tsupko-2015}, here we have written the electron number density in the equatorial plane as $N=N(r)$ where $r=\sqrt{b^2+z^2}$, with $z$ denoting the coordinate along the axis parallel to the incoming ray. For more details about calculations with formulas (\ref{alpha-refr-R}) and (\ref{alpha-refr-b}) see, e.g., \cite{BK-Tsupko-2015, BK-Tsupko-2023-time-delay}. Formula (\ref{alpha-refr-b}) was also derived in \cite{BK-Tsupko-2009} and \cite{BK-Tsupko-2010}. Calculations of refractive deflection for power-law plasma distributions can be found, e.g., in \cite{Bliokh-Minakov-1989, BK-Tsupko-2009, BK-Tsupko-2010, BK-Tsupko-2015, BK-Tsupko-2023-time-delay}, see also \cite{Crisnejo-Gallo-Jusufi-2019}.

Finally, we obtain the total deflection angle as a function of the impact parameter $b$:
\begin{equation} \label{total-deflection-b}
\hat{\alpha}(b) = \frac{4m}{b} + \frac{15 \pi}{4} \frac{m^2}{b^2} \mp \frac{4ma}{b^2} + \hat{\alpha}_{\mathrm{refr}}(b) \, ,
\end{equation}
where the first three terms describe vacuum gravitational deflection up to second order in $m$ and first order in $a$, and the last term is the refractive deflection, $\hat{\alpha}_{\mathrm{refr}}(b)$, given by Eq.(\ref{alpha-refr-b}). Note that in the following, when writing $\hat{\alpha}(...)$, we will always mean the functional dependence on $b$ as in (\ref{total-deflection-b}), rather than on $R$ as in (\ref{total-deflection-R}).

With our definition of $\hat{\alpha}$, the gravitational (i.e., vacuum) deflection angle is positive, as it is usually considered. The refractive deflection $\hat{\alpha}_{\mathrm{refr}}(b)$ can have either sign, depending on the specific plasma distribution. In view of applications to astrophysics it is realistic to assume that the plasma density decreases with the radial coordinate ($dN/dr<0$). In this case, in the weak-deflection approximation the refractive deflection is negative: ($\hat{\alpha}_{\mathrm{refr}}(b)<0$), see \cite{BK-Tsupko-2010}. The total angle $\hat{\alpha}$ can be either positive and negative, remaining small. Recall that a positive value of $\hat{\alpha}$ means that the light ray is bent towards the center.

For the sake of completeness we conclude this Section with giving some further references where higher-order terms for vacuum deflection in the equatorial plane of a Kerr black hole can be found: \cite{Iyer-Hansen-2009a, Aazami-Keeton-2011b, Chakraborty-Sen-2015}. Exact expressions for the vacuum deflection angle in the Kerr spacetime are examined in \cite{Iyer-Hansen-2009b, Congdon-2022-kerr-bending}. The vacuum deflection angle by a Kerr black hole when the incoming trajectory is parallel to the rotation axis was found already in \cite{Skrotskii-1957}, see also \cite{Dymnikova-1986} and \cite{Epstein-Shapiro-1980}. Gravitational lensing by a Kerr black hole in vacuum (derivation and solution of lens equation, calculation of image properties) was also investigated in the following works: \cite{Ibanez-1983, Dymnikova-1984, Dymnikova-1986, Bray-1986-rotating-lens, Bliokh-Minakov-1989, Rauch-Blandford-1994, Glicenstein-1999, Asada-2003-rotating-lens, Sereno-2002, Sereno-2006-kerr-weak, Werner-Petters-2007, Chen-Chen-Wang-2007, Bozza-2008-kerr-caustics, Aazami-Keeton-2011a, Aazami-Keeton-2011b}.

\section{Influence of the black-hole spin and of a plasma on image positions}
\label{sec:images}

In this Section, we consider primary and secondary images of point sources produced by gravitational lensing of a Kerr black hole surrounded by a plasma. We study the influence of the black-hole spin and of the plasma on the position of these images, assuming that the observer and the light sources are in the equatorial plane and working in the regime of weak deflection.

It should be noted that, in general, both the black-hole rotation and the presence of a plasma can lead to the appearance of additional images. Recall that in the Schwarzschild spacetime without a plasma, in the weak deflection approximation there are two images, a primary one where the azimuthal coordinate sweeps out an angle between $0$ and $\pi$ and a secondary one where the azimuthal angle sweeps out an angle between $\pi$ and $2 \pi$. In the following we stick with the weak-deflection approximation and we calculate the correction to the position of these two images as they are produced by the spin of the black hole and by the plasma density, both of which are assumed to be small. A similar approach was used in  \cite{BK-Tsupko-2023-time-delay} where linear plasma corrections to the time delay of primary and secondary images in the Schwarzschild spacetime were calculated. We mention that the appearance of additional images due to the presence of a plasma is crucial for microlensing phenomena, see \cite{Tsupko-BK-2020-microlensing, Sun-Er-Tsupko-2023} for the influence of a plasma on microlensing.

In this section we use the standard lens equation which is based on the assumptions that light sources are in a plane, that the gravitational field is weak and that light rays are bent only in a lens plane by a deflection angle that is small. As shown in Fig.~\ref{fig-kerr-lens}, we assume that both the observer and the light source are in the equatorial plane of the Kerr metric; then the lens equation relates the angular position of a source $\beta$ to the angular positions $\theta$ of images. (The angle $\beta$ should not be confused with the function $\beta_p$ used earlier in Sec.\ref{sec:weak}.) Different forms of a lens equation, based on various approximations, are known in the literature \cite{GL1, GL2, Frittelli-2000, Virbhadra-2000, Perlick-2004-exact-equation, Bozza-Sereno-2006, Bozza-2008-lens-equations, Aazami-Keeton-2011a, Aazami-Keeton-2011b}.

In the presence of a plasma the lens equation can be applied in the usual form only if we assume that the plasma density falls off to zero if infinity is approached. As the Kerr metric is asymptotically flat, this implies that the light rays are asymptotically straight lines. If both the observer and the source are situated very far from the black hole, and if all angles involved are small, $\beta, \theta, \hat{\alpha} \ll 1$, the lens equation can then be written as \cite{GL1, GL2}
\begin{equation} \label{le}
\beta = \theta - \frac{D_{ds}}{D_s} \hat{\alpha} \, .
\end{equation}
In this lens equation, it is supposed that the angle $\beta$ is positive, while the angle $\theta$ can be either positive or negative.

\begin{figure}
\includegraphics[width=0.95\columnwidth]{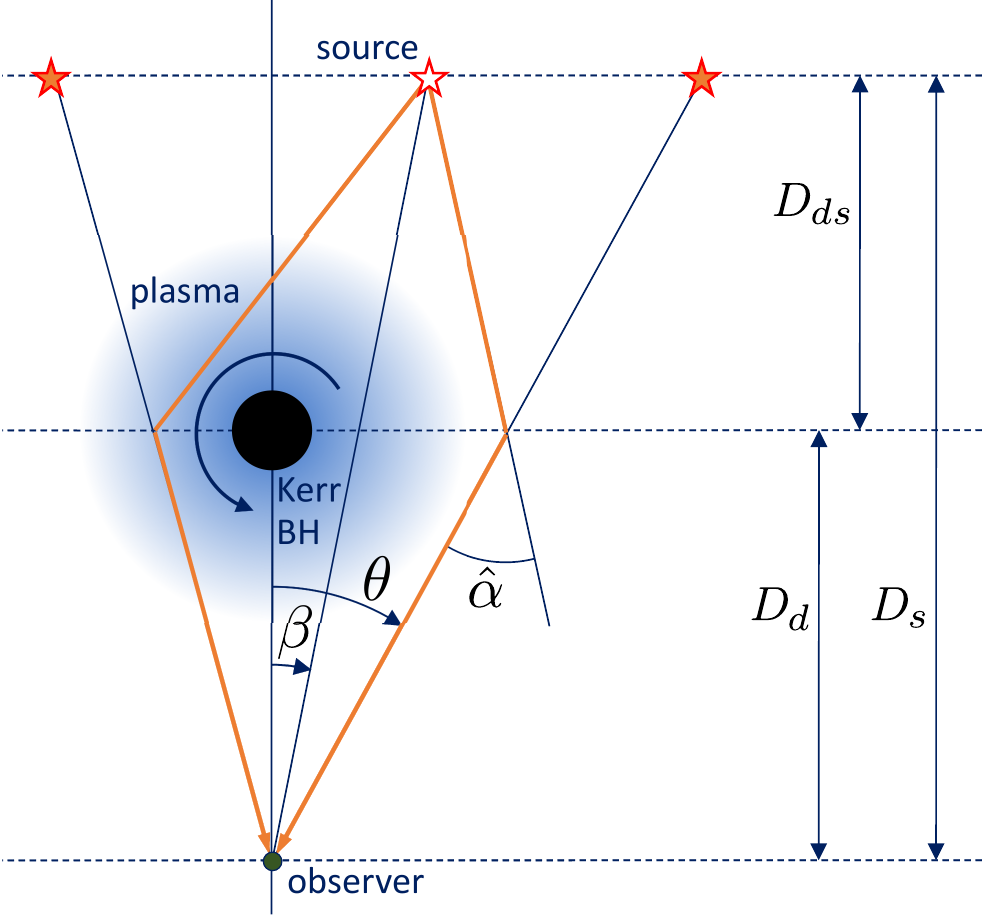}
\caption{Lensing in the equatorial plane of a Kerr black hole surrounded by plasma in the weak-deflection approximation. The a angular position of the source is given by $\beta>0$. The angular positions of the images are given by $\theta$. The primary image is the image located on the same side of the lens as the source ($\theta_+>0$), and the secondary image is the image on the opposite side ($\theta_-<0$). The light ray in the left part of the picture co-rotates with the black hole, whereas the one in the right part counter-rotates. The angles $\beta$, $\theta$, $\hat{\alpha}$ are assumed to be small. The source and the observer are assumed to be far from the black hole, $D_i \gg m$.}
\label{fig-kerr-lens}
\end{figure}

Note that in Eqs. (\ref{alpha-refr-b}) and (\ref{total-deflection-b}) the argument $b$ is considered as positive. Therefore, in order to allow negative $\theta$, the lens equation for the spherically symmetric case can be rewritten as (e.g., Eq.(17) of \cite{Suyu-2012}, Eq.(13) on p.104 of \cite{GL2}, Eq.(19) of \cite{BK-Tsupko-2023-time-delay}):
\begin{equation} \label{lens-eq-02}
\beta = \theta - \alpha(|\theta|) \frac{\theta}{|\theta|} \, ,
\end{equation}
where 
\begin{equation}
\alpha(|\theta|) = \frac{D_{ds}}{D_s} \, \hat{\alpha}(|D_d \theta|) \, .
\label{eq:alphahatalpha}
\end{equation}
Here we have used the approximate relation $b=D_d |\theta|$, which is valid for small $\theta$. The angle $\hat{\alpha}$ can be decomposed into a contribution from the gravitational field in vacuum and a contribution from the plasma which was denoted $\hat{\alpha} _{\mathrm{refr}}$ in the preceding section. For practical purposes, the main interest is in density profiles that are everywhere decreasing with the radial coordinate, which implies $\hat{\alpha}_{\mathrm{refr}}<0$. Since it is more convenient to work with positive values, we introduce a function $B_\omega(|\theta|)$ by (cf.\cite{BK-Tsupko-2023-time-delay, Tsupko-BK-2020-microlensing})
\begin{equation}
\frac{D_{ds}}{D_s} \, \hat{\alpha}_{\mathrm{refr}} = - B_\omega(|\theta|) \, .
\end{equation}
We mean here that after substituting $b=D_d |\theta|$ into Eq.(\ref{alpha-refr-b}) we obtain some function of $|\theta|$ which, after multiplication by distances and changing sign, is then denoted $B_\omega(|\theta|)$. The function $B_\omega(|\theta|)$ is positive for density profiles decreasing with the radial coordinate.

Now we use Eq.(\ref{total-deflection-b}) with $b=D_d |\theta|$ and find:
\begin{equation} \label{alpha-normalized}
\alpha(|\theta|) = \frac{\theta_E^2}{|\theta|} + \frac{\theta_E^2}{|\theta|^2} \varepsilon_m    \mp \frac{\theta_E^2}{|\theta|^2} \varepsilon_a  -  \varepsilon_\omega B_\omega(|\theta|)   \, .
\end{equation}
Here $\theta_E$ is the vacuum angular Einstein radius,
\begin{equation}
\theta_E^2 = 4m \frac{D_{ds}}{D_d D_s} \, ,
\end{equation}
and we have introduced the parameters
\begin{equation}
\varepsilon_m = \frac{15 \pi}{16}  \frac{m}{D_d}  , \quad \varepsilon_a = \frac{a}{D_d} \, .
\end{equation}
For consistency, on the right-hand side of (\ref{alpha-normalized}) we have also introduced a book-keeping parameter $\varepsilon_\omega$ which will be set equal to unity after all equations have been linearized with respect to it.

With (\ref{alpha-normalized}), the lens equation (\ref{lens-eq-02}) finally becomes:
\begin{equation} \label{lens-eq-03}
    \beta = \theta - \left[ \frac{\theta_E^2}{|\theta|} + \frac{\theta_E^2}{|\theta|^2} \varepsilon_m    \mp \frac{\theta_E^2}{|\theta|^2} \varepsilon_a  -  \varepsilon_\omega B_\omega(|\theta|) \right]  \frac{\theta}{|\theta|} \, .
\end{equation}
We remind the reader that the upper sign in Eq. (\ref{alpha-normalized}) and in the lens equation (\ref{lens-eq-03}) corresponds to co-rotation, and the lower sign corresponds to counter-rotation.

Now we will solve the lens equation (\ref{lens-eq-03}) perturbatively, in the approximation that $\varepsilon_m$, $\varepsilon_a$ and $\varepsilon_\omega$ are so small that only terms of linear order with respect to these parameters have to be taken into account. Then the solution of zeroth order represents the Schwarzschild case with the deflection term $\theta_E^2/\theta$, whereas the second-order Schwarzschild deflection, the deflection due to the black-hole rotation and the plasma refraction are linear perturbations independent of each other. We will consider the case shown in Fig.~\ref{fig-kerr-lens}, where the primary image is formed by a ray that is counter-rotating with respect to the black hole, and the secondary image is formed by a ray that is co-rotating. As throughout this paper, we assume that $a \ge 0$.
If the black hole rotates in opposite direction compared to the picture, we will just need to change a sign in front of all terms with $a$. Perturbative solutions of the lens equation were also determined for strong lens systems with a singular isothermal ellipsoid model of the lens in \cite{Crisnejo-Gallo-2023-perturbative} and for a singular isothermal sphere model in \cite{BK-Tsupko-2023-time-delay}.

Let us begin with considering the primary image, for which the corresponding ray is counter-rotating and $\theta > 0$, $|\theta| = \theta$, see Fig.~\ref{fig-kerr-lens}. Therefore the lens equation (\ref{lens-eq-03}) takes form:
\begin{equation} \label{eq-primary}
\beta = \theta - \frac{\theta_E^2}{\theta} - \frac{\theta_E^2}{\theta^2} \varepsilon_m  - \frac{\theta_E^2}{\theta^2} \varepsilon_a  +  \varepsilon_\omega B_\omega(\theta)  \, .
\end{equation}
We will seek the solution in the following form:
\begin{equation} \label{theta_seek}
\theta = \theta^{(0)} + \varepsilon_m \theta^{(m)} + \varepsilon_a \theta^{(a)} + \varepsilon_\omega \theta^{(\omega)} \, .
\end{equation}
Here $\theta^{(0)}$, $\theta^{(m)}$, $\theta^{(a)}$ and $\theta^{(\omega)}$ are unknown functions: $\theta^{(0)}$ is the zeroth-order term, and the other terms are first-order corrections associated with the second-order Schwarzschild deflection, the black hole spin and the plasma, respectively.

Substituting expression (\ref{theta_seek}) into equation (\ref{eq-primary}) and keeping the terms linear in $\varepsilon^{(m)}$, $\varepsilon^{(a)}$ and $\varepsilon^{(\omega)}$, we obtain:

\begin{equation} \label{eq-primary-expanded}
\beta = \theta^{(0)} + \varepsilon_m \theta^{(m)} + \varepsilon_a \theta^{(a)} + \varepsilon_\omega \theta^{(\omega)}  -
\end{equation}
\[
- \frac{\theta_E^2}{\theta^{(0)}} + \varepsilon_m \theta^{(m)} \frac{\theta_E^2}{(\theta^{(0)})^2} + \varepsilon_a \theta^{(a)} \frac{\theta_E^2}{(\theta^{(0)})^2} + \varepsilon_\omega \theta^{(\omega)} \frac{\theta_E^2}{(\theta^{(0)})^2} -  
\]
\[
- \frac{\theta_E^2}{(\theta^{(0)})^2} \varepsilon_m  - \frac{\theta_E^2}{(\theta^{(0)})^2} \varepsilon_a  +  \varepsilon_\omega B(\theta^{(0)}) \, .
\]
The zeroth-order equation
\begin{equation}
\beta = \theta^{(0)} - \frac{\theta_E^2}{\theta^{(0)}}
\end{equation}
for positive $\theta^{(0)}$ gives the position of the primary image for the point-mass Schwarzschild lens:
\begin{equation} \label{zero-primary}
\theta^{(0)}_+ = \frac{\beta}{2} + \frac{1}{2} \sqrt{\beta^2 + 4 \theta_E^2 } \, .
\end{equation}
Combining the terms containing $\varepsilon_m$, we obtain the first-order equation for the Schwarzchild correction $\theta^{(m)}$,
\begin{equation}
0 = \theta^{(m)} + \theta^{(m)} \frac{\theta_E^2}{(\theta^{(0)})^2} - \frac{\theta_E^2}{(\theta^{(0)})^2} \, , 
\end{equation}
and its solution
\begin{equation}
\theta^{(m)} = \frac{\theta_E^2}{\theta_E^2 + (\theta^{(0)})^2} \, .
\end{equation}
Analogously, we obtain:
\begin{equation}
\theta^{(a)} = \frac{\theta_E^2}{\theta_E^2 + (\theta^{(0)})^2} \, ,
\end{equation}
\begin{equation}
\theta^{(\omega)} = - \frac{(\theta^{(0)})^2  B_\omega(\theta^{(0)})}{\theta_E^2 + (\theta^{(0)})^2} \, .
\end{equation}
Finally, we get the following expression for the position of the primary image:
\begin{equation}
\theta_+ = \theta^{(0)}_+   +  \frac{15 \pi}{16}  \frac{m}{D_d}  \frac{\theta_E^2}{\theta_E^2 + (\theta^{(0)}_+)^2} + 
\end{equation}
\[
+  \frac{a}{D_d}  \frac{\theta_E^2}{\theta_E^2 + (\theta^{(0)}_+)^2} - \frac{(\theta^{(0)}_+)^2  B_\omega(\theta^{(0)}_+)}{\theta_E^2 + (\theta^{(0)}_+)^2} \, ,
\]
where $\theta^{(0)}_+$ is defined in Eq.(\ref{zero-primary}).

Let us also present this solution as an explicit function of $\beta$. After some algebra we get:
\begin{equation} \label{theta-plus}
\theta_+ = \theta^{(0)}_+  \, +  \frac{15 \pi}{32}  \frac{m}{D_d}  w^{(-)}(\beta) \, +
\end{equation}
\[
+ \, \frac{a}{2D_d} w^{(-)}(\beta)    \,   -  \frac{1}{2} w^{(+)}(\beta)   B_\omega(\theta^{(0)}_+) \, ,
\]
where we have introduced the function
\begin{equation} \label{eq:w-beta}
w^{(\pm)}(\beta) = 1 \pm \frac{\beta}{\sqrt{\beta^2 + 4 \theta_E^2 } }  \,  .
\end{equation}
Note that the plus-minus sign in $w^{(\pm)}(\beta)$ has is unrelated to the plus-minus sign that distinguishes between primary and secondary images.\\

Now let us consider the secondary image. Here we have $\theta < 0$, $|\theta| = - \theta$ and the light ray is co-rotating (Fig.~\ref{fig-kerr-lens}). Therefore the lens equation (\ref{lens-eq-03}) takes form:
\begin{equation} \label{eq-secondary}
\beta = \theta - \frac{\theta_E^2}{\theta} + \frac{\theta_E^2}{\theta^2} \varepsilon_m  - \frac{\theta_E^2}{\theta^2} \varepsilon_a  -  \varepsilon_\omega B_\omega(-\theta) \, .
\end{equation}
Analogously to the primary image considered above, we obtain the solution:
\begin{equation}
\theta_- = \theta^{(0)}_-   -  \frac{15 \pi}{16}  \frac{m}{D_d}  \frac{\theta_E^2}{\theta_E^2 + (\theta^{(0)}_-)^2} + 
\end{equation}
\[
+  \frac{a}{D_d}  \frac{\theta_E^2}{\theta_E^2 + (\theta^{(0)}_-)^2} + \frac{(\theta^{(0)}_-)^2  B_\omega(-\theta^{(0)}_-)}{\theta_E^2 + (\theta^{(0)}_-)^2} \, ,
\]
where $\theta^{(0)}_-$ is the zeroth-order solution for the secondary image,
\begin{equation} \label{zero-secondary}
\theta^{(0)}_- = \frac{\beta}{2} - \frac{1}{2} \sqrt{\beta^2 + 4 \theta_E^2 } < 0 \, .
\end{equation}
or, as an explicit function of $\beta$:
\begin{equation} \label{theta-minus}
\theta_- = \theta^{(0)}_-  \, -  \frac{15 \pi}{32}  \frac{m}{D_d}  w^{(+)}(\beta) \, +
\end{equation}
\[
+  \frac{a}{2D_d} w^{(+)}(\beta)    \,   + \frac{1}{2} w^{(-)}(\beta)   B_\omega(-\theta^{(0)}_-) \, ,
\]
where $w^{(\pm)}(\beta)$ is defined in (\ref{eq:w-beta}).

The second-order Schwarzschild corrections in (\ref{theta-plus}),(\ref{theta-minus}) agree with the results of  \cite{Ebina-2000, Keeton-Petters-2005, Keeton-Petters-2006}. Both second-order terms (second-order Schwarzschild and first-order Kerr) agree with \cite{Aazami-Keeton-2011b} (see formula (32) there and non-numbered formulas further), see also \cite{Sereno-2002}, \cite{Sereno-2006-kerr-weak}.\\

\begin{figure*}
\includegraphics[width=0.90\textwidth]{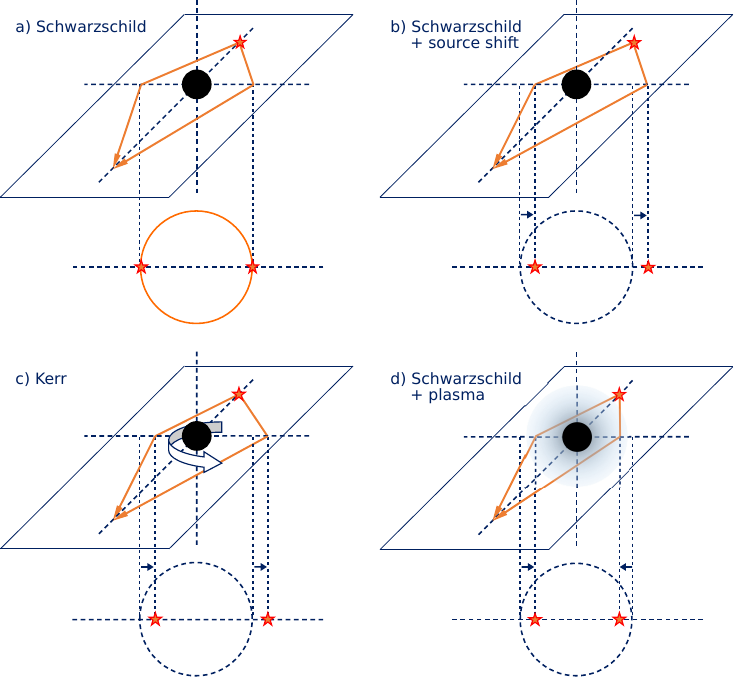}
\caption{Comparison of different effects on the position of primary and secondary images. The case of a Schwarzschild lens in vacuum with perfect alignment of source, lens and observer is shown in panel a). In this case the observer would see an Einstein ring. The equatorial plane with two rays is chosen arbitrarily in this picture. A shift $\beta$ of the source relative to the observer-lens line, see panel b), leads to a breaking of the Einstein ring. Now there are two images which are shifted in the direction of the source position. To within first order in $\beta$, the displacements of images are equal, i.e., the angular separation between the two images is the same as the diameter of the Einstein ring in panel a). The rotation of the black hole, see panel c), also leads to a shift of both images. To within first order with respect to the spin parameter $a$, both images are shifted equally. Thus, also in this case the angular separation between the two images equals the diameter of the Einstein ring in panel a). By contrast, the presence of a plasma with decreasing density profile around a Schwarzschild black hole, see panel d), leads to a shift of both images towards the lens. Therefore, the angular separation between them is smaller than the diameter of the Einstein ring in panel a), already in the first order of the plasma correction. See more details in the text.}
\label{fig-kerr-plasma}
\end{figure*}

The expressions (\ref{theta-plus}) and (\ref{theta-minus}) for the image positions can be significantly simplified if we additionally assume that $\beta \ll \theta_E$. In this case, the angular positions of the two images are close to an Einstein ring. The zeroth-order solutions are:
\begin{equation} \label{theta-zero-approx}
\theta^{(0)}_+ = \theta_E + \frac{\beta}{2} \, , \quad \theta^{(0)}_- = - \theta_E + \frac{\beta}{2} < 0 \, .
\end{equation}
If $\beta$ is so small that it can be neglected in all the linear corrections derived above, we obtain for the primary image:
\begin{equation} \label{theta-plus-approx}
\theta_+ = \theta_E + \frac{\beta}{2}   +  \frac{15 \pi}{32}  \frac{m}{D_d} + \frac{1}{2} \frac{a}{D_d}  - \frac{1}{2} B_\omega(\theta_E) \, ,
\end{equation}
and for the secondary image:
\begin{equation} \label{theta-minus-approx}
\theta_- = - \theta_E + \frac{\beta}{2}   -  \frac{15 \pi}{32}  \frac{m}{D_d} + \frac{1}{2} \frac{a}{D_d}  + \frac{1}{2} B_\omega(\theta_E) \, .
\end{equation}
The angular separation between the two images is:
\begin{equation} \label{angular-separation}
\Delta \theta = \theta_+ - \theta_- = 2 \theta_E  + \frac{15 \pi}{16}\frac{m}{D_d}    - B_\omega(\theta_E) 
\end{equation}
The first two terms on the right-hand side give the separation of the two images for a Schwarzschild lens with second-order corrections taken into account. The last term is due to the presence of the plasma. We see that, in the approximation of Eq. (\ref{angular-separation}), neither a shift of the source nor an increase of the black-hole spin leads to a change of the angular separation: in first order, the corresponding terms cancel each other. Only the presence of the plasma leads to a change, namely to a decrease, of the angular separation.

On the basis of Eqs.(\ref{theta-zero-approx}), (\ref{theta-plus-approx}), (\ref{theta-minus-approx}) and (\ref{angular-separation}) we can compare the influence of different corrections on the image positions, see Fig.~\ref{fig-kerr-plasma}.

In Fig.~\ref{fig-kerr-plasma}a), there is a Schwarzschild lens in the case of perfect alignment of source, lens and observer ($\beta=0$). In this case  the observer will see a circular image known as an Einstein ring. For purposes of vizualization, we have arbitrarily chosen the equatorial plane of this black hole, and have shown only two light rays in this plane which give two points of the Einstein ring. Now we discuss how the positions of these two images are changed due to different effects.

The second-order Schwarzschild correction makes the photon deflection angle bigger, i.e., it leads to a slight increase of the size of the Einstein ring. This effect is not drawn separately in the picture.

In Fig.~\ref{fig-kerr-plasma}b), we show the Schwarzschild lens in the case that the source is shifted relative to the observer-lens line ($\beta>0$). In this case, the rotational symmetry is broken, so the observer does not see a ring but rather two images. They are shifted in the direction of the source position. As follows from Eqs.(\ref{theta-zero-approx}), (\ref{theta-plus-approx}) and (\ref{theta-minus-approx}), in the first order of angular shift $\beta$ of the source, the displacements of images are equal by absolute value which means that the angular separation between the two images remains the same as in the case of perfect alignment shown in panel a).

Fig.~\ref{fig-kerr-plasma}c) shows a Kerr lens in the case of perfect alignment of source, lens and observer with the source and the observer in the equatorial plane. Similarly to panel b), the Einstein ring of panel a) breaks into two images, but now because of the rotation of the lens. In the case of co-rotation (left ray), the rotation of the black hole reduces the gravitational deflection angle, i.e., a ray with the same impact parameter is deflected by a smaller angle as compared to the Schwarzschild case. Therefore, the light ray is forced to go closer to the black hole in order to reach to the observer at the same location, i.e., it must have a smaller impact parameter than the corresponding ray in the Schwarzschild case. Therefore, the left image is shifted towards the lens. In the case of counter-rotation (right ray), the rotation of the black hole increases the angle of deflection. As a result, the ray forming the image passes the lens at a greater distance, i.e., its impact parameter is bigger than the corresponding ray in the Schwarzschild case. Therefore, the right image is shifted away from the lens.

As follows from Eqs.(\ref{theta-zero-approx}), (\ref{theta-plus-approx}) and (\ref{theta-minus-approx}), to within first order with respect to the spin parameter $a$, both images are shifted by the same amount. Thus, the angular separation between the two images in this case also does not change, i.e. it remains equal to the diameter of the Einstein ring in the Schwarzschild case.

It is easy to see that the influence of the spin parameter $a$ on the image positions is equivalent (in the first-order approximation ) to the angular shift of the source from the observer-lens line by the angle $\beta = a/D_d$, see \cite{Dymnikova-1986} and references there, and also \cite{Sereno-2006-kerr-weak}. Therefore, even if the observer knows the position of the lens and knows that the shift of both images takes place, the shift due to the black-hole rotation is indistinguishable from the shift of images due to a shift of the source.

In Fig.~\ref{fig-kerr-plasma}d), we show the Schwarzschild black hole surrounded by a plasma with a density that is decreasing with increase of the radial coordinate. The presence of the plasma reduces the total angle of deflection. Therefore, to reach the observer, both light rays in the picture come closer to the lens in comparison to the vacuum case. As a result, the observed images are closer to the lens, and the angular separation between them is smaller. The magnitude of this effect depends, of course, on the ratio of the plasma density and the photon frequency.

\section{Conclusions}

(i) In this paper, light propagation in a non-magnetized pressure-free plasma in the domain of outer communication of a Kerr black hole is considered, which is a continuation of our previous study (Paper I \cite{Perlick-Tsupko-2017}). We assume throughout that the plasma density depends on the coordinates $r$ and $\vartheta$ according to (\ref{eq:sepcon}) which allows for separability of the Hamilton-Jacobi equation for the light rays, i.e., for the existence of a Carter constant. The necessary and sufficient condition for separability was derived in Paper I. Here we have focused on the analysis of different types of orbits and have found several phenomena which do not exist in the vacuum case.

(ii) Starting from the equations of motion for a light ray in the Kerr metric in the presence of a plasma that satisfies the separability condition, Eqs.(\ref{eq:dott}), (\ref{eq:dotphi}), (\ref{eq:dottheta}) and (\ref{eq:dotr}), we have investigated spherical and conical orbits in Section \ref{sec:spherical-conical}. It is revealed that there can exist two spherical light rays travelling through the same point, but that this can happen only inside the ergoregion. We have also shown that, as for the spherical light rays, there are at most two conical light rays through a point outside the equatorial plane. In the same Section, we give the condition for the existence of light rays in the equatorial plane, see Eq. (\ref{eq:plsymmetry}), and their equations of motion, see Eqs. (\ref{eq:dott2}), (\ref{eq:dotphi2}) and (\ref{eq:dotr-new}).

(iii) In Section \ref{sec:circular} we have derived the necessary and sufficient conditions for the existence of circular light orbits on and off the equatorial plane, see Eqs. (\ref{eq:circgen}) and (\ref{eq:circoff}). In contrast to the vacuum case, where circular light rays off the equatorial plane do not exist in the domain of outer communication, in a plasma with a $\vartheta$-dependent density such orbits are possible. This was already briefly mentioned in Paper I but is discussed in more detail, with examples, here. Moreover, in the same section the equation for circular light rays in the equatorial plane is written in a compact form, see Eq.(\ref{eq:circ-final}), and several particular cases are discussed.

(iv) In Section \ref{sec:shadow} we have continued our discussion of the influence of a plasma on the shadow of a Kerr black hole, which was the main topic of Paper I, in particular the appearance of ``fishtails'' in the boundary curve of the shadow, see Fig.6 in Paper I. We have found that the actual appearance of the shadow strongly depends on the location of light sources and that in no case the observer will actually see the fishtails in the sky.

(v) We have given an exact formula for the deflection angle of a light ray in the equatorial plane in Section \ref{sec:exact}, see Eqs.(\ref{eq:halpha}), (\ref{eq:deff}), and compared it to an earlier formula derived by Perlick \cite{Perlick-2000}. The main advantage of the new formula is that it is suitable for all rays that come in from infinity and return to infinity, including those that enter into the ergoregion. We have also discussed the notion of impact parameter in relation to other constants of motion in the presence of a plasma. In particular, the expressions \ref{impact-par-plasma-03} and \ref{impact-par-plasma-04} are found.

(vi) For several examples we have graphically investigated the exact dependence of the deflection angle on the distance of closest approach, and then also on the impact parameter (Section \ref{sec:exact}). Remarkably, it is found that the dependence can be non-monotonic, see Figs. \ref{fig:Schwdef3} and  \ref{fig:Schwdef4}.

(vii) We have studied and compared the effects of the black-hole spin and of the plasma on gravitational lensing in the weak-deflection approximation, see Sections \ref{sec:weak} and \ref{sec:images}. We have calculated the deflection angle of light rays in the equatorial plane of the Kerr metric with arbitrary value of the spin parameter and in the presence of an arbitrary spherically-symmetric plasma distribution. The lens equation is solved perturbatively and angular positions of primary and secondary images are found, see Eqs.(\ref{theta-plus}) and (\ref{theta-minus}). Whereas the presence of the black hole spin does not lead, in first order of $a$, to a change of the angular separation between two images in comparison with the Schwarzschild case, the plasma presence does, already in the first order of the function $\omega_p^2/\omega_0^2$, see Fig.\ref{fig-kerr-plasma}.

\section*{Acknowledgements}

OYuT thanks Emanuel Gallo for a motivating discussion which led to the writing of Section \ref{sec:shadow}. The work of OYuT is partially supported by a Humboldt Research Fellowship for experienced researchers; OYuT thanks Claus Lämmerzahl for great hospitality. OYuT is also thankful to DAAD (Deutscher Akademischer Austauschdienst) for support of his visit to ZARM, Bremen University, in December 2018 -- January 2019, where this work was started. OYuT especially thanks G.S. Bisnovatyi-Kogan for support of all scientific initiatives.


%

\end{document}